\documentclass[
  aps,
  prb,
  reprint,
  superscriptaddress,
  longbibliography,
  showkeys,
  citeautoscript,
]{revtex4-2}
\usepackage[intlimits]{amsmath}
\usepackage{amssymb}
\usepackage{amsthm}
\usepackage{booktabs}
\usepackage{braket}
\usepackage{color}
\usepackage{graphicx}
            
\usepackage[colorlinks=true,
            linkcolor=black,
            citecolor=black,
            filecolor=black,
            urlcolor=black
            ]{hyperref}
\usepackage[utf8]{inputenc}
\usepackage[version=4]{mhchem}
\usepackage{multirow}
\usepackage{soul}
\usepackage{tabularx}
\usepackage{units}
\usepackage[dvipsnames]{xcolor}

\usepackage{dcolumn}
\usepackage{booktabs}
\usepackage{miller}

\newcommand{\dtu}{
    Department of Electrical and Photonics Engineering, Technical University of Denmark,
    2800 Kgs. Lyngby, Denmark
}

\newcommand{\nbi}{
    Center for Quantum Devices, Niels Bohr Institute, University of Copenhagen, 2100
København Ø, Denmark
}

\newcommand{\tum}{
Walter Schottky Institute, Technical University of Munich, 85748 Garching, Germany
}

\makeatletter
\newcommand\emailx[1]{%
\move@AF%
\def\@affil{{\normalfont\,#1\strut}{}}%
}%

\begin{document}

\title{Toward single-photon detection with superconducting niobium diselenide nanowires}

\author{Pietro Metuh}
\affiliation{\dtu}

\author{Athanasios Paralikis}
\affiliation{\dtu}

\author{Paweł Wyborski}
\affiliation{\dtu}

\author{Sherwan Jamo}
\affiliation{\nbi}

\author{Alessandro Palermo}
\affiliation{\tum}

\author{Lucio Zugliani}
\affiliation{\tum}

\author{Matteo Barbone}
\affiliation{\tum}

\author{Kai Müller}
\affiliation{\tum}

\author{Niels Gregersen}
\affiliation{\dtu}

\author{Saulius Vaitiek\.{e}nas}
\affiliation{\nbi}

\author{Jonathan Finley}
\affiliation{\tum}

\author{Battulga Munkhbat}
\email{bamunk@dtu.dk}
\affiliation{\dtu}

\begin{abstract}

We present superconducting nanowire single-photon detectors (SNSPDs) based on few-layer NbSe$_2$ fully encapsulated with hexagonal boron nitride (hBN), demonstrating single-photon sensitivity. Our fabrication process preserves the superconducting properties of NbSe$_2$ in nanowires, as confirmed by low-temperature transport measurements that show a critical temperature of $T_c \approx 6.5$~K, comparable to the reported values for unpatterned sheets, and it maintains a contact resistance of $\sim 50 \, \Omega$ at $T = 4$~K.  Meandered NbSe$_2$ nanowires exhibit a responsivity of up to $4.9 \times 10^4$ V/W over a spectral range of 650-1550 nm in a closed-cycle cryostat at 4~K, outperforming planar and short-wire devices. The devices achieve a $1/e$ recovery time of $\tau = (135 \pm 36)$ ns, system timing jitter of $j_\text{sys} = (1103 \pm 7)$ ps, and detection efficiency of $\sim 0.01\%$ at $0.95I_c$, with a linear increase in detection probability confirming the single-photon operation. Furthermore, measurements under attenuated pulsed laser (1 MHz) indicate a success rate of up to $33\%$ in detecting individual optical pulses, establishing the platform as a promising candidate for developing efficient single-photon detectors.

\end{abstract}

\maketitle

\section{Introduction}

In photonic quantum technologies, the qubit is encoded in a single photon, enabling secure communication protocols for quantum key distribution \cite{zahidy2024quantum} and the manipulation of pure, indistinguishable, quantum states \cite{maring2024versatile}. The key components for these technologies are the single-photon sources to generate the qubits, the photonic circuits to harness them, and the detectors to read them at the end of the transmission or computation \cite{flamini2018photonic}. However, efficiently detecting single photons remains a significant challenge, as only some classes of photodetectors are sensitive to the level of a single photon.  
Single-photon avalanche diodes (SPADs) and photomultiplier tubes (PMTs) emerged first as detectors of single photons, but tradeoffs between high detection efficiency, rapid response, and low-noise operation have limited their efficacy \cite{hadfield2009single}. 
Conversely, superconducting nanowire single-photon detectors (SNSPDs) excel in terms of recovery time, timing jitter, and detection efficiency \cite{esmaeil2021superconducting}, with recent demonstrations even achieving photon number resolution (up to four photons) without signal demultiplexing \cite{cahall2017multi}. 
While conventional SNSPDs fabricated from materials such as niobium-based superconductors have achieved impressive figures of merit and have been successfully integrated in photonic chips for single photon detection \cite{reithmaier2013chip}, their deposition techniques involve complex processes and strict controls. For instance, the quality of ultrathin (3-6 nm) films of niobium nitride (NbN) and niobium titanium nitride (NbTiN), typically grown with magnetron sputtering or molecular beam epitaxy, is highly dependent on several parameters, such as ion bombardment, gas pressure, and substrate temperature \cite{cucciniello2022superconducting, miki2009superconducting}. 
In addition, the decreasing critical temperature with thinner films, which is linked to structural degradation on both surfaces \cite{ilin2008ultra, licata2022correlation}, may constrain their on-chip integration and scalability.

\begin{figure*}[t!]
    \centering
    \includegraphics[width=\linewidth]{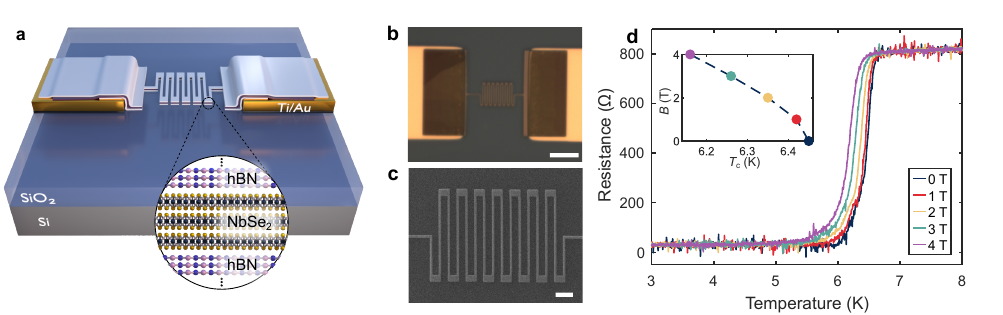}
    \caption{Fabrication of NbSe$_2$ nanowires. (a) Illustration of a fabricated NbSe$_2$ superconducting nanowire with top and bottom hBN encapsulation; the inset shows the crystalline structure of the van der Waals materials. (b) Optical microscopy image of an encapsulated 120 nm-wide nanowire with a resist mask patterned with a meander geometry; the scale bar is 10 \textmu m. (c) Scanning electron microscopy image of the nanowire; the scale bar is 1 \textmu m. (d) Five-point moving-average magnetoresistance of an encapsulated 500 nm-wide straight wire as a function of temperature, with the inset showing the field dependence of the critical temperature.}
    \label{fig:one}
\end{figure*}

As an alternative, two-dimensional superconductors offer a promising platform, thanks to a reduced electronic heat capacity and electron-phonon coupling \cite{wang2022emerging}. For instance, magic-angle twisted bilayer graphene has recently shown exceptional sensitivity to single photons \cite{di2024infrared}. However, such detectors are currently limited by their longer recovery time ($\sim 5$ ms) and lower absorption \cite{montblanch2023layered}, which are not compatible with quantum communications and computing. 
2H-niobium diselenide (2H-NbSe$_2$) provides another option for a superconducting platform. As a van der Waals material belonging to the class of transition-metal dichalcogenides (TMDs), NbSe$_2$ is widely available as a bulk crystal, from which single- or few-layer flakes can be exfoliated, simplifying the fabrication process while still enabling the use of a crystalline superconductor \cite{staley2009electric, mills2014single}. In addition, this transferring technique simplifies integrating a detector on pre-fabricated photonic waveguides -- an essential attribute for scalable quantum photonic circuits -- and it is compatible with other TMD-based photonic components \cite{paralikis2024tailoring, munkhbat2023nanostructured}.
The critical temperature of NbSe$_2$ depends on the number of layers (from $T_c \approx 3$ K for 1L-NbSe$_2$ to $T_c \approx 7$~K for the bulk material) \cite{frindt1972superconductivity, xi2016ising} and, while it is lower than NbN \cite{tian2020improved, licata2022correlation, cukauskas1983effects}, it remains compatible with standard helium-based cryogenic systems for bilayer NbSe$_2$ ($T_c \approx 5$ K) or thicker flakes.  
Additionally, NbSe$_2$ can benefit from encapsulation with hexagonal boron nitride (hBN) \cite{shein2024fundamental}, which has been extensively used for thermal management or creating a stable charge environment in other van der Waals materials. Moreover, capping the NbSe$_2$ samples with hBN immediately after transferring NbSe$_2$ might be crucial to suppress oxidation, which is the main contributor to degradation in NbSe$_2$ \cite{orchin2021two, el2013superconductivity} and other niobium-based films \cite{cucciniello2022superconducting}.

To date, multiphoton detection with NbSe$_2$ flakes has been shown \cite{orchin2019niobium}, but detection of singlephotons has not been demonstrated yet. However, to enable sensitivity to single photons, it is also essential to pattern the flakes and reduce the dimensionality of NbSe$_2$. Previous works have shown the compatibility of NbSe$_2$ with standard patterning methods and reported prototypes of bolometers \cite{mills2014single, shein2024towards, jayanand2024optically, jiang2023fast}. However, the impact of the fabrication techniques on the transport properties of NbSe$_2$ and on its sensitivity to light remains uncertain, and a NbSe$_2$ single-photon detector has not been demonstrated yet \cite{montblanch2023layered}.

In this work, we present superconducting nanowire photodetectors based on few-layer NbSe$_2$ that are fully encapsulated with hBN, demonstrating single-photon sensitivity. Our fabrication process is validated through cryogenic transport measurements, which confirm that even narrow wires retain their superconducting properties—showing no significant alteration in critical temperature or magnetoresistance compared to unpatterned flakes—and maintain a low contact resistance ($\sim 50 \, \Omega$ at $T = 4$~K).
Moreover, we investigated the optical sensitivity of meanderered  NbSe$_2$ nanowires by comparing them with unpatterned flakes and short straight wires. Under illumination over a broad spectral range (650–1550~nm) in a closed-cycle optical cryostat (T = 4 K), our devices achieve high responsivity up to $4.9 \times 10^4$~V/W.
We also estimated $1/e$ recovery time  of $\tau = (135 \pm 36)$~ns, and measured system timing jitter ($j_\text{sys} = (1103 \pm 7)$~ps) and detection efficiency ($\sim 0.01\%$ at $0.95I_c$), which are mainly limited by the electronics of the setup. Furthermore, our measurements under attenuated pulsed laser (1 MHz) indicate a success rate of up to $33\%$ in detecting individual optical pulses. Finally, the linear increase in detection probability with the number of photons provides clear evidence of single-photon sensitivity. These results position our platform as a promising candidate for developing ultrathin, crystalline SNSPDs for future photonic quantum technologies.

\bigskip
\section{Nanofabrication and magnetoresistance measurements}
To fabricate NbSe$_2$ photodetectors, we begin by preparing the substrate, thermally growing 150 nm of SiO$_2$ on silicon and subsequently patterning electrodes (5 nm Ti / 50 nm Au) with UV lithography and electron-beam evaporation. Thin flakes ($\lesssim 5$ nm) of 2H-NbSe$_2$ are exfoliated from a bulk crystal (HQ Graphene) with blue tape, filtered by size, homogeneity, optical contrast and color with an optical microscope, and transferred onto the substrate. 
The exact number of layers can be determined with Raman spectroscopy or atomic force microscopy \cite{orchin2019niobium, staley2009electric}. However, to minimize the exposure of the flakes to the environment, we immediately proceed to cap the flake with either hBN or electron-beam resist.
In hBN-encapsulated samples, preselected flakes of hBN are transferred both before and after the transfer of the NbSe$_2$; an illustration of this process is reported in Supporting Figure 1. The bottom flake of hBN is interposed between NbSe$_2$ and the SiO$_2$ film, but does not entirely shield the NbSe$_2$ flake from the gold electrodes to maintain electrical contact. The upper hBN layer is then transferred to cover the entire surface of the flake that is to be patterned.
To avoid further degradation of the NbSe$_2$, the samples are immediately cleaned with acetone, isopropyl alcohol, and de-ionized water, spin-coated with a negative electron-beam resist (AR-N 7520), and exposed with an electron-beam writer (acceleration voltage of 30 or 100 kV, depending on the sample). The resist is developed in AR 300-47 for 90 s. Fluorine-based inductively coupled plasma reactive-ion etching (ICP-RIE) is carried out to transfer the pattern onto the flakes. Different combinations of gases have been tested. The best results were obtained with a combination of CF$_4$, which is more aggressive yet less polymerizing than CHF$_3$, and argon (25 W coil power, 25 W platen power, 2.5 mTorr). 
An alternative etching method based on anisotropic wet chemistry is also possible \cite{munkhbat2023nanostructured} and was tested by producing a similar etching mask; fabrication details are reported in Supporting Figure 2.

Figure \ref{fig:one}a shows the resulting device after RIE. When characterizing the sample, the resist mask is not removed from the upper surface of the NbSe$_2$ structure for increased protection against oxidation. Samples that have not been encapsulated in hBN but that maintain a resist mask after fabrication will be referred to as resist-encapsulated (RE) samples. Conversely, samples whose NbSe$_2$ flake has been encapsulated with lower and upper hBN layers will be called fully encapsulated (FE). Finally, flakes without encapsulation are referred to as not encapsulated (NE). Optical microscopy (OM) and scanning electron microscopy (SEM) images showing an example of the resulting nanowire (patterned into a meander geometry) are given in Figure \ref{fig:one}b,c.

Previous works have suggested that patterned ultrathin NbSe$_2$ might not be conductive \cite{mills2014single, el2013superconductivity}. Therefore, we first investigate whether the superconducting properties of the flake are preserved after the fabrication steps. Two fully encapsulated straight nanowires were measured using standard low-frequency lock-in techniques in a kiutra L-Type Rapid cryostat. The first sample was patterned into a single 500 nm-wide wire and was probed with a three- or four-point method, while the second sample comprised two parallel, 500 nm-wide wires connected in the middle and was probed with a two-point method; both devices showed comparable results.
In the first wire sample (Figure \ref{fig:one}d) we find a critical temperature (defined as the point at which the sample resistance is half of the normal-state resistance) of $T_c = 6.45$~K at zero magnetic field, which is consistent with previously reported values for few-layer NbSe$_2$ flakes \cite{frindt1972superconductivity, xi2016ising}. 
The field-dependent critical temperature in the inset shows a typical parabolic decrease with increasing magnetic field \cite{Tinkham1996}. Optical images of both samples and additional transport measurements are reported in Supporting Figure 3. 
Incidentally, it is worth pointing out that the contact and wiring resistance contribution ($53 \, \Omega$) measured in the second sample is smaller than previous works with nanopatterned NbSe$_2$ \cite{mills2014single}. Other samples in this work have shown similar values of background resistance, ranging from $32 \, \Omega$ to $128 \, \Omega$. The reduced contact resistance may arise from transferring NbSe$_2$ after, rather than before, the deposition of Ti and Au electrodes.

\begin{figure}[t]
    \centering
    \includegraphics[width=\columnwidth]{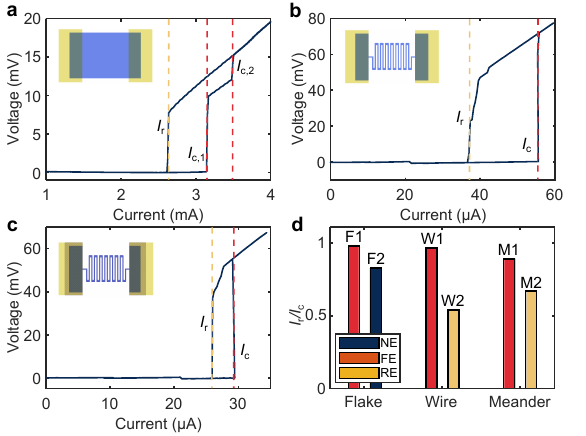}
    \caption{Electrical characterization of NbSe$_2$ structures in a closed-cycle vacuum cryostat ($T = 3.9$ K). (a-c) I-V curves of (a) an unpatterned NbSe$_2$ flake with variable thickness, (b) A patterned, resist-encapsulated NbSe$_2$ meander wire, and (c) A patterned, fully encapsulated NbSe$_2$ meander wire, with their critical currents $I_c$ and retrapping currents $I_r$. The contact and wiring resistance was substracted for clarity. (d) Ratio of retrapping current to critical current for two wide flakes (F1 and F2), two short-wire samples (W1 and W2), and two meander samples (M1 and M2), characterized at the base temperature. NE = no encapsulation; FE = full encapsulation; RE = resist encapsulation.}
    \label{fig:two}
\end{figure}

\bigskip
\section{Electrical measurements in a closed-cycle vacuum cryostat}
All other measurements on these and other samples were carried out in an optical, closed-cycle vacuum cryostat (attocube attoDRY800xs) with a base temperature of $T \approx 3.9$ K. To evaluate the electrical properties of the device, the latter is connected to a voltage source in series with a 100 k$\Omega$ resistor. While this cryostat does not have the same degree of accuracy for setting the sample temperature due to the limited cooling power from the cold head underneath the sample, the samples display superconductivity, which can be suppressed by applying a sizeable current. Due to the temperature gradient between the sensor and the sample surface and the absence of a helium atmosphere in the cryostat chamber, the actual temperature on the sample surface is slightly ($\lesssim 1$~K) higher than the recorded value (Supporting Figure 4).

I-V curves (shown only for positive current values but otherwise symmetrical for the negative current axis) of fabricated devices were obtained by linearly increasing and decreasing the bias current and by subsequently subtracting a linear resistance contribution arising from the contacts, wiring, and 100-k$\Omega$ resistor. First, we report a curve for an unpatterned, non-encapsulated NbSe$_2$ flake (F1) that was transferred on contact pads and loaded immediately into the cryostat (Figure \ref{fig:two}a). When sweeping the current left to right, two main transitions were identified, with critical current values of $I_{\text{c},1} = 3.16$~mA and $I_{\text{c},2} = 3.49$~mA. This is likely due to the inhomogeneity in thickness of the unpatterned flake, which has both a thinner region (where the critical current is supposedly lower) and a thicker region (where the necessary current-induced heat to break superconductivity is higher). As reported in other works \cite{orchin2019niobium, shein2024fundamental}, a large hysteresis is observed when decreasing the current. In contrast to the critical current, a single retrapping transition is observed. The retrapping current, i.e. the value at which the superconductive state is restored, is $I_r = 2.63$~mA. Compared to the critical current values, the retrapping current is $I_r = 0.83 I_{\text{c},1} = 0.75I_{\text{c},2}$.

Patterning the flake into a meander wire (130 nm in width, 500 nm in pitch, 340 nm in bend width, and 90 \textmu m in total length) decreases considerably the critical current for NbSe$_2$ flakes of similar thickness due to the much smaller cross-section of the wire. Figure \ref{fig:two}b reports the I-V curve for a RE meander wire, which is even more hysteretic than the plain flake of Figure \ref{fig:two}a. However, if the flake is fully encapsulated (Figure \ref{fig:two}c), the hysteresis of a similar meander (120 nm in width, 500 nm in pitch, 390 nm in bend width, and 90 \textmu m in length) seems to decrease considerably. This behavior can be attributed to the additional hBN layers, whose outstanding in-plane room temperature thermal conductivity might contribute to dissipating the current-induced heat more efficiently than the polymer resist and the silicon oxide substrate \cite{jo2013thermal, sichel1976heat}.

The ratio between the retrapping and critical current $I_r/I_c$ (which quantifies how little a device is affected by hysteresis) of various samples (two wide flakes, two flakes patterned into straight wires, and two patterned into longer meanders) is reported in the histogram of Figure \ref{fig:two}d. Remarkably, all the samples with the least hysteresis are fully encapsulated, independently of the dimensionality. This further hints at the role of the hBN layers in dissipating charge-induced heat, which reduces the possibility of the device latching when operated as a photodetector. 
Moreover, thinner wires generally show a higher degree of hysteresis than shorter ones when fully encapsulated, with a wider encapsulated flake showing nearly no hysteresis at the base temperature of $T = 3.9$~K ($I_r = 0.99 I_c$), the thicker wires characterized in Figure \ref{fig:one}d (W1) having slightly more hysteresis ($I_r = 0.97 I_c$), and the meander wire shown in Figure \ref{fig:one}b,c (M1) having the largest hysteresis among the fully encapsulated samples. OM images and I-V curves for W1, W2, F1, and F2 are shown in Supporting Figures S5 and S6.
As previously reported in other works \cite{li2011retrapping, sahu2009individual, orchin2019niobium}, increasing the sample temperature induces a decrease in hysteresis due to the critical current decreasing at a faster rate than the retrapping current (Supporting Figure 7).

\begin{figure}[t]
    \centering
    \includegraphics[width=\columnwidth]{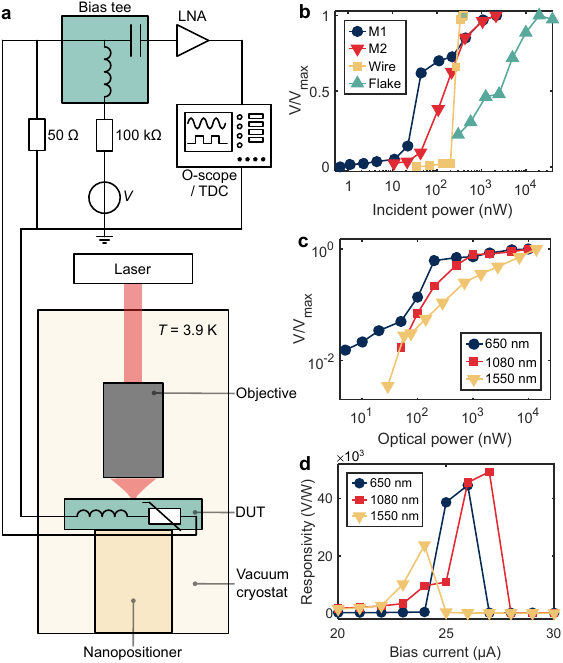}
    \caption{Optical characterization of NbSe$_2$ photodetectors. (a) Illustration of the experimental setup to optically characterize the NbSe$_2$ photodetectors. The device under test (DUT), represented by a variable resistor and an inductor, is cooled by a cold plate in the closed-cycle cryostat and irradiated with a focused laser beam. A bias voltage $V$ is applied to the circuit, which is completed by a high-load resistor ($R_s = 100 \text{ k}\Omega$) and, when using an oscilloscope or a time-to-digital converter, a parallel resistor ($R_p = 50 \, \Omega$); the detector response is either measured at the voltage source (when using CW lasers) or at the oscilloscope or time-to-digital converter after an amplifier. (b) Comparison of the voltage response of four samples under laser irradiation, normalized by the amplitude as the devices saturate ($ T = 5$~K). The meanders (M1 and M2) and the wire (W1) are irradiated with a 650 nm CW laser, and the flake (F1) is irradiated with a 1080 nm pulsed laser. The incident power is normalized by the irradiated area of the device. (c) Voltage response of sample M1 at three CW laser wavelengths ($ T = 5$~K). Optical response of the unpatterned flake irradiated by a 1064 nm pulsed laser with an 8 kHz repetition rate. (d) Voltage responsivity of sample M1 against the bias current at 220 nW of incident power for three CW laser wavelengths ($ T = 4$~K).}
    \label{fig:three}
\end{figure}

\bigskip
\section{Optical characterization and responsivity}
To assess the optical properties of the device, the electrical circuit is extended by including a bias tee (Picosecond Pulse Labs 5575A) to block the source DC component, an amplifier at the capacitor port (Mini-Circuits ZFL-500LN+), and an oscilloscope (Rohde \& Schwarz RTB2004) or time-to-digital converter (TDC; Swabian Instruments Time Tagger Ultra). The device is illuminated with laser sources at various wavelengths (650, 1064, 1080, 1550 nm), whose collimated beam enters the optical window of the cryostat and is focused by a low-temperature objective (attocube LT-APO/Telecom). 

First, the optical sensitivity was tested by using a calibrated laser source (continuous-wave 650 nm for the meander and wire samples and 1064 nm for the unpatterned NbSe$_2$ flake) and measuring the generated amplitude difference at the voltage source, normalized by the maximum value. An optical chopper system modulates the CW laser and decouples the optical response from the oscillations caused by temperature fluctuations in the closed-loop vacuum cryostat, whose temperature was set to $T = 5$~K to reduce latching. For the same reason, the bias current was chosen to be near the critical current but separate enough to avoid transitions to the normal state between the cooling cycles of the cryostat. The effect of the cooling cycle on the voltage response is reported in Supporting Figure 8.
The thinner, fully encapsulated meander M1 shows sensitivity to the laser light down to the power value of 1 nW (measured before the cryostat window and normalized by the overlap between the device cross-section and the laser beam). As the dimensionality of the devices increases, so does the minimum incident power needed to observe a minimum amplitude response. In all cases, the amplitude response eventually saturates as the excessive laser energy drives the sample out of the superconducting state.
It is worth mentioning that the dataset for the unpatterned flake was taken with a pulsed laser (8 kHz, 1064 nm). However, limited measurements with the same 650 nm laser showed an optical response at similar orders of magnitude, considerably higher than the patterned samples.

\begin{figure*}[t]
    \centering
    \includegraphics[width=0.9\textwidth]{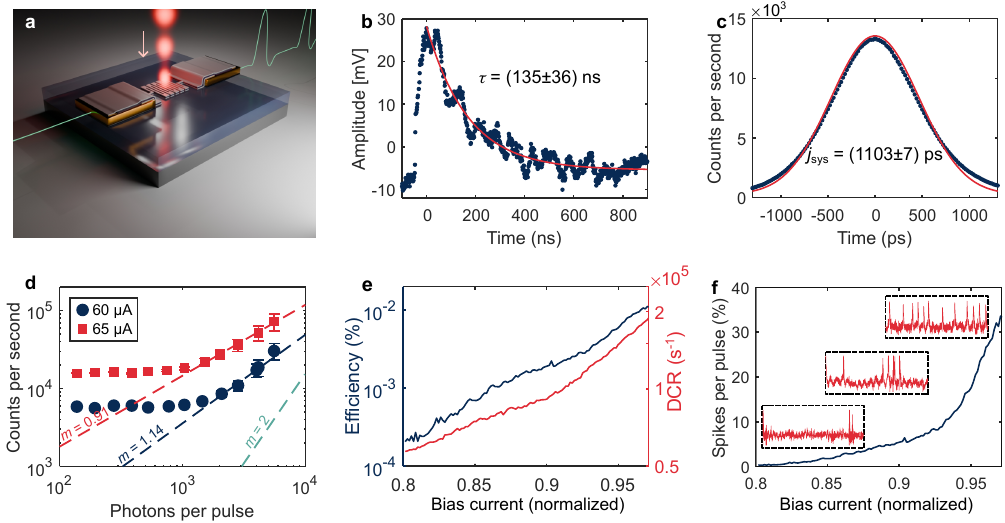}
    \caption{Toward NbSe$_2$ SNSPDs. (a) Illustration of the detection of attenuated laser pulses on the encapsulated meander sample, resulting in voltage spikes. (b) A typical pulse measured with an oscilloscope. The recovery time, measured as the $1/e$ decay constant, is calculated by averaging 138 peaks. (c) Measurement of the timing jitter $j_\text{sys}$ of the entire system in the experimental setup. (d) Detected counts at the TDC against the number of photons (femtosecond, 1100 nm) in each pulse for three current bias values. The lines are obtained by fitting the points above the noise count level. The error bars show one standard deviation. (e) Detection efficiency and dark count rate against the normalized bias current (normalized by the critical current). (f) Detected laser pulses against the normalized bias current.}
    \label{fig:four}
\end{figure*}

For other measurements, we report the result of the more sensitive sample (M1), which, thanks to the meander geometry, enhances the absorption cross-section of the wire when illuminated by a focused laser beam \cite{natarajan2012superconducting}. Figure \ref{fig:three}c shows the dependence of the voltage response on the laser wavelength (650, 1080, and 1550 nm) at different optical power values (measured before the cryostat window). While the response for 1080 nm and 1550 nm follows a similar trend and shows a higher sensitivity at lower optical power values, the one at 650 nm shows an initially lower sensitivity at low power values, and a second, faster sensitivity regime at approximately the same power regime as the other wavelength. The initial lower sensitivity might be attributed to higher photon absorption by the Si/SiO$_2$ substrate at 650 nm, which might slightly increase the temperature on the sample surface. 
Nonetheless, the sample shows sensitivity to all three wavelengths, hinting at the possibility of broadband operation for NbSe$_2$ photodetectors.

Finally, the voltage responsivity, namely the ratio between the generated amplitude difference and the incident power, was measured at different bias current values (Figure \ref{fig:three}d) at $T = 4$~K. Here, the optical power was fixed to $P_\text o = 1$~\textmu W (corresponding to $\lessapprox 220$~nW of incident power) for all three wavelengths. The responsivity initially increases with the bias current, reaching a maximum of $ R = 4.9 \times 10^4$~V/W at 650 nm, and decreases sharply as the current approaches $I_\text c$. A similar trend is observed at 1080 nm, whose responsivity peaks at $ R = 4.4 \times 10^4$~V/W. The maximum responsivity measured at 1550 nm is $R = 2.3 \times 10^4$~V/W. We partly attribute the shift to the left of the responsivity peak at 1550 nm to a delay of two weeks between the measurement of this dataset compared to the other two, which has been shown to cause reductions in the critical current \cite{orchin2019niobium}.

\bigskip
\section{Toward single photon detection}
At this stage, we evaluate the detection dynamics with an attenuated laser and either an oscilloscope or a TDC. The device, biased with the circuit shown in Figure \ref{fig:three}a, shows its response by switching from the superconducting to the normal state upon detection of the attenuated pulses, thereby generating a spike in voltage that the oscilloscope or the TDC can read. An illustration of the mechanism for the detector is shown in Figure \ref{fig:four}a.
To compensate for the well-known latching behavior that affects niobium-based superconducting detectors \cite{annunziata2010reset, orchin2019niobium}, a resistor of 50 $\Omega$ is inserted in parallel with the device. This creates a low-resistance path when the device is triggered by a pulse and operates in the normal state, such that current-induced heating is dissipated and superconductivity is restored. The modified I-V curve of the meander sample after inserting the parallel resistor is reported in Supporting Figure 9.

While short (100 ps or shorter) optical pulses could not be detected by the planar flake F1, longer pulses ($ \lessapprox 2$~ns) could. An oscilloscope trace of such detections is shown in Supporting Figure 10.
On the other hand, a response could be resolved with the patterned meander M1 under irradiation of attenuated pulses from pico- and femtosecond lasers. A typical peak response under a pulsed laser (1080 nm, 1 MHz repetition rate, $\sim 100$ ps pulse length) is shown in Figure \ref{fig:four}b, where an inverse exponential curve is fitted. Analyzing a trace of detection events (Supporting Figure 11) with 138 peaks revealed a recovery time, which can be estimated from the decay dynamics and is often quoted as the $1/e$ decay time for asymmetric pulses \cite{morozov2021superconducting}, of $\tau = (135 \pm 36)$ ns. An alternative definition (i.e., the time required for the pulse to decrease from 90\% to 10\% \cite{yin2024heat}) gives a recovery time of $ \tau' = (297 \pm 79)$ ns.
The timing jitter of the entire system was also measured by correlating pulse events with the arrival times of the detection events at the TDC, and its distribution with a Gaussian fitting curve is shown in Figure \ref{fig:four}c. The obtained jitter value from the FWHM of the fitted Gaussian curve is $ j_\text{sys} = (1103 \pm 7)$ ps. We attribute this large value to the poor signal-to-noise ratio in our experimental setup, which is caused by the long ($\gtrapprox 3$ m) room-temperature DC line between the cryostat and the electronics, which acts as a transmission line for electrical noise and suffers from crosstalk with the other lines. At this stage, we could not isolate and measure the timing jitter contribution of the cryostat interface; however, the significant noise in the system is unrelated to the NbSe$_2$ detectors. Therefore, neglecting the smaller jitter values of the other electronic components ($< 50$ ps) and the picosecond laser ($< 100$ ps), the timing jitter contribution of the NbSe$_2$ sample is likely lower than this reported value.

By adjusting the power of an attenuated laser at 1100 nm (with a repetition rate of 80 MHz and pulse width of $\sim 100$ fs), we measured the count rate at the TDC as a function of the number of photons per pulse at three values of bias current in logarithmic scale. Similar to the timing jitter, the considerable noise in the system severely affects this measurement, which shows a very high dark count rate (DCR), as evidenced by the flat line for a low number of photons ($\lesssim 1000$) per pulse. This rate, however, should not only be interpreted as the detection of thermal photons but mainly as a symptom of the fluctuations in the electronics of the system.
As the number of photons increases, the signal counts overcome the noise counts, and a linear increase in the count rate with the number of photons can be observed. We fit the data points above the noise level with a line and evaluate the slope $m$ for the three bias current values. The lower values of  $60$ and 65 \textmu A are characterized by a near-unity slope (1.14 and 0.91, respectively) in a logarithmic scale, which demonstrates single-photon sensitivity of the detectors \cite{gol2001picosecond, fitzpatrick2010superconducting, di2024infrared}. The sub-unity slope (0.49) at 70 \textmu A suggests that the detector is already near saturation above the noise level.

We then measure the count rate with and without a laser source by sweeping the bias current to evaluate the DCR and estimate detection efficiency using the calibrated source method \cite{hadfield2009single}. The attenuated laser at 1080 nm and 1 MHz is used. Due to the high DCR (which includes the noise counts) shown on the right-hand axis of Figure \ref{fig:four}e that must be subtracted to the detected counts with the pulsed laser, the resulting system detection efficiency at $0.95 I_c$ is $\eta \sim 0.01\%$ at 1080 nm.
We also show the percentage of the detected laser pulses resulting in a detection event as a function of the bias current with a repetition rate of 1 MHz (Figure \ref{fig:four}f). Here, the pulse detection rate is 25\% at $0.95I_c$, which must be regarded as a lower bound, given the subtracted number of dark counts. Oscilloscope traces at different bias current values (shown in the insets) illustrate the signal spikes matching the laser repetition rate.

\bigskip
\section{Discussion}

We presented a fabrication workflow for patterning NbSe$_2$ nanowires with upper and lower hBN encapsulation and showed that their critical temperature and its magnetic field dependence are similar to those reported for unpatterned NbSe$_2$ flakes. The observed values for the contact resistance indicate good contact between NbSe$_2$ and gold, even without high-temperature annealing, which might be due to the quick transferring of NbSe$_2$ flakes after exfoliation, thus limiting the oxidation of the portion of the flake in contact with gold and avoiding intermediate fabrication steps that could impact the interface between the two materials.
We then showed a correlation between the degree of hysteresis in the superconductivity transition and the encapsulation method and dimension of the device. By decreasing the gap between retrapping and critical current, lower and upper hBN encapsulation was shown to be beneficial. These findings add NbSe$_2$ to the list of van der Waals materials benefitting integration with hexagonal boron nitride \cite{yankowitz2019van, daveau2020spectral, piccinini2024high}.

We also showed the optical sensitivity of NbSe$_2$ detectors as a function of operating wavelength, optical power, and bias current. The high responsivity measured at 650, 1080, and 1550 nm in a meander demonstrates the suitability of NbSe$_2$ for photodetection in the visible and telecom range. Interestingly, we observed hints of decreases in critical current after two weeks, which have previously been linked to aging of the sample \cite{orchin2019niobium}. Given the encapsulation on both upper and lower surface of the sample, it is possible that this is phenomenon is related to the lateral edge of the sample, but a systematic investigation needs to be carried out to confirm this.
Furthermore, the key performance parameters related to SNSPDs have been estimated for this system. However, when determining the system timing jitter, dark count rate and the detection efficiency of the devices, several limitations were encountered due to the electrical noise, which was introduced by three main factors. The first (and less severe) was related to the fluctuation of the sample temperature throughout the cooling cycles and to the general operating temperature ($T = 3.9$ K) being relatively close to the critical temperature of the sample measured in the system ($T = 5.6$ K). 
Secondly, the experiment was carried out in an optically unshielded cryostat, which introduces several dark counts caused by the laboratory background noise \cite{gol2001picosecond}. 
The last source of noise was considerably more severe and is attributed to the DC line used in the characterization. As previously mentioned, this line increases considerably the length between the sample and the (room temperature) amplifier, and it does not generally support low-noise, high-frequency operation. Partial improvements could be obtained by short-circuiting all the unused lines during operation. However, as shown in Supporting Figure 12, the noise level remains high, even when no bias is applied or when connected to another cryostat line, excluding the sample as the source of this electrical noise. 
Despite these limitations, we presented evidence of sensitivity to single photons by registering a linear increase of the count rate with the number of incident photons.
Moreover, we showed the possibility of early-stage characterization of SNSPD properties in a single-stage closed-cycle vacuum cryostat without dedicated electronic radio-frequency (RF) lines, which might decrease the operational cost for single-photon detectors and increase their flexibility, pushing their adoption across other fields \cite{kramer2023helium, morozov2021superconducting, gemmell2017miniaturized}.

Future work should be directed toward optimizing the detection efficiency by improving the design of the nanowire. A decrease in cross section of the wire by at least a factor three are in reach by making thinner devices or by patterning narrower wires with a higher filling factor. In addition, the efficiency can be improved by choosing a more appropriate substrate to enhance absorption in the NbSe$_2$ layers, and even integrating the device in an optical cavity \cite{esmaeil2021superconducting, zugliani2023tailoring}. A more careful choice of the shunt resistor might also decrease the detector recovery time \cite{dauler2014review}. At the same time, decoupling the device performance from the limitations of the experimental system should be prioritized to establish NbSe$_2$ as a competitive van der Waals platform for detecting single photons.
Given the flexibility in fabrication that NbSe$_2$ offers, the platform could be particularly suitable for integration with other structures for on-chip single-photon processing.  

\section{Conclusion}
In summary, we have demonstrated superconducting nanowire single-photon detectors based on few-layer NbSe$_2$ fully encapsulated with hBN that operate at the single-photon level. Our fabrication process preserves the intrinsic superconducting properties of NbSe$_2$, as verified by cryogenic transport measurements and low contact resistance ($\sim 50 \, \Omega$). Our experimental results reveals high responsivity (up to $4.9 \times 10^4$ V/W  over 650–1550 nm) and promising device metrics, including a 1/e recovery time of $\tau = (135 \pm 36)$ ns), timing jitter of $j_\text{sys} = (1103 \pm 7)$ ps, and detection efficiency of $\sim 0.01\%$ at $I = 0.95I_c$. Furthermore, the linear response and a $33\%$  success rate for detecting low-power pulsed signals confirm promising single-photon operation. These findings make our NbSe$_2$ nanowire based SNSPD platform a promising candidate for scalable quantum photonic circuits, with the potential for further optimization to enhance performance and integration in advanced quantum communication and computing systems.

\section*{Data availability}

The data supporting the findings of this study are available within the main manuscript and its supplementary information files. Additional data are available from the corresponding authors upon reasonable request.

\section*{Acknowledgements}

The authors acknowledge support from the European Research Council (ERC-StG ``TuneTMD", grant no. 101076437) and Villum Fonden (project no. VIL53033 and no. VIL53097). The authors also acknowledge the European Research Council (ERC-CoG ``Unity", grant no. 865230) and the cleanroom facilities at the Danish National Centre for Nano Fabrication and Characterization (DTU Nanolab).  JJF gratefully acknowledges the German Science Foundation for financial support through the DFG Clusters of Excellence - Munich Center for Quantum Science and Technology (MCQST, EXC2111) and e-conversion (EXC2089), as well as grants INST95-1496-1, INST 95/1719-1, FI947-8, and SPP2244 (FI 947/7-2).  In addition, JJF acknowledges the State of Bavaria for funding via Munich Quantum Valley programme via the iQSense project.  

\section*{Associated content}
\textbf{Supporting information.} 

Illustration of the fabrication process; optical microscopy images of other processed samples; additional temperature-dependent resistance measurements and I-V curves; effect of the closed-loop cryostat cooling cycles in the optical responsivity; effect of the electrical noise on timing jitter and spurious dark counts.

\section*{Author contributions}

PM and AtP fabricated the samples. 
SA and SV characterized the critical temperature of the samples and its dependence on the magnetic field. 
PM performed the imaging and the electrical and optical characterization with support from PW. 
PM performed the data analysis and processing of all the data. 
AlP, LZ, and JJF measured the low-temperature transport on preliminary devices and contributed to the characterization setup.
BM supervised and coordinated the project. 
PM, AtP, PW, NG, and BM wrote the manuscript with the input of all co-authors.

\section*{Competing Interests}
The authors declare no competing interest.

\appendix

\bibliography{biblio}

\providecommand{\noopsort}[1]{}\providecommand{\singleletter}[1]{#1}%
\begin{thebibliography}{47}%
\makeatletter
\providecommand \@ifxundefined [1]{%
 \@ifx{#1\undefined}
}%
\providecommand \@ifnum [1]{%
 \ifnum #1\expandafter \@firstoftwo
 \else \expandafter \@secondoftwo
 \fi
}%
\providecommand \@ifx [1]{%
 \ifx #1\expandafter \@firstoftwo
 \else \expandafter \@secondoftwo
 \fi
}%
\providecommand \natexlab [1]{#1}%
\providecommand \enquote  [1]{``#1''}%
\providecommand \bibnamefont  [1]{#1}%
\providecommand \bibfnamefont [1]{#1}%
\providecommand \citenamefont [1]{#1}%
\providecommand \href@noop [0]{\@secondoftwo}%
\providecommand \href [0]{\begingroup \@sanitize@url \@href}%
\providecommand \@href[1]{\@@startlink{#1}\@@href}%
\providecommand \@@href[1]{\endgroup#1\@@endlink}%
\providecommand \@sanitize@url [0]{\catcode `\\12\catcode `\$12\catcode `\&12\catcode `\#12\catcode `\^12\catcode `\_12\catcode `\%12\relax}%
\providecommand \@@startlink[1]{}%
\providecommand \@@endlink[0]{}%
\providecommand \url  [0]{\begingroup\@sanitize@url \@url }%
\providecommand \@url [1]{\endgroup\@href {#1}{\urlprefix }}%
\providecommand \urlprefix  [0]{URL }%
\providecommand \Eprint [0]{\href }%
\providecommand \doibase [0]{https://doi.org/}%
\providecommand \selectlanguage [0]{\@gobble}%
\providecommand \bibinfo  [0]{\@secondoftwo}%
\providecommand \bibfield  [0]{\@secondoftwo}%
\providecommand \translation [1]{[#1]}%
\providecommand \BibitemOpen [0]{}%
\providecommand \bibitemStop [0]{}%
\providecommand \bibitemNoStop [0]{.\EOS\space}%
\providecommand \EOS [0]{\spacefactor3000\relax}%
\providecommand \BibitemShut  [1]{\csname bibitem#1\endcsname}%
\let\auto@bib@innerbib\@empty
\bibitem [{\citenamefont {Zahidy}\ \emph {et~al.}(2024)\citenamefont {Zahidy}, \citenamefont {Mikkelsen}, \citenamefont {M{\"u}ller}, \citenamefont {Da~Lio}, \citenamefont {Krehbiel}, \citenamefont {Wang}, \citenamefont {Bart}, \citenamefont {Wieck}, \citenamefont {Ludwig}, \citenamefont {Galili} \emph {et~al.}}]{zahidy2024quantum}%
  \BibitemOpen
  \bibfield  {author} {\bibinfo {author} {\bibfnamefont {M.}~\bibnamefont {Zahidy}}, \bibinfo {author} {\bibfnamefont {M.~T.}\ \bibnamefont {Mikkelsen}}, \bibinfo {author} {\bibfnamefont {R.}~\bibnamefont {M{\"u}ller}}, \bibinfo {author} {\bibfnamefont {B.}~\bibnamefont {Da~Lio}}, \bibinfo {author} {\bibfnamefont {M.}~\bibnamefont {Krehbiel}}, \bibinfo {author} {\bibfnamefont {Y.}~\bibnamefont {Wang}}, \bibinfo {author} {\bibfnamefont {N.}~\bibnamefont {Bart}}, \bibinfo {author} {\bibfnamefont {A.~D.}\ \bibnamefont {Wieck}}, \bibinfo {author} {\bibfnamefont {A.}~\bibnamefont {Ludwig}}, \bibinfo {author} {\bibfnamefont {M.}~\bibnamefont {Galili}}, \emph {et~al.},\ }\href@noop {} {\bibfield  {journal} {\bibinfo  {journal} {npj Quantum Information}\ }\textbf {\bibinfo {volume} {10}},\ \bibinfo {pages} {2} (\bibinfo {year} {2024})}\BibitemShut {NoStop}%
\bibitem [{\citenamefont {Maring}\ \emph {et~al.}(2024)\citenamefont {Maring}, \citenamefont {Fyrillas}, \citenamefont {Pont}, \citenamefont {Ivanov}, \citenamefont {Stepanov}, \citenamefont {Margaria}, \citenamefont {Hease}, \citenamefont {Pishchagin}, \citenamefont {Lema{\^\i}tre}, \citenamefont {Sagnes} \emph {et~al.}}]{maring2024versatile}%
  \BibitemOpen
  \bibfield  {author} {\bibinfo {author} {\bibfnamefont {N.}~\bibnamefont {Maring}}, \bibinfo {author} {\bibfnamefont {A.}~\bibnamefont {Fyrillas}}, \bibinfo {author} {\bibfnamefont {M.}~\bibnamefont {Pont}}, \bibinfo {author} {\bibfnamefont {E.}~\bibnamefont {Ivanov}}, \bibinfo {author} {\bibfnamefont {P.}~\bibnamefont {Stepanov}}, \bibinfo {author} {\bibfnamefont {N.}~\bibnamefont {Margaria}}, \bibinfo {author} {\bibfnamefont {W.}~\bibnamefont {Hease}}, \bibinfo {author} {\bibfnamefont {A.}~\bibnamefont {Pishchagin}}, \bibinfo {author} {\bibfnamefont {A.}~\bibnamefont {Lema{\^\i}tre}}, \bibinfo {author} {\bibfnamefont {I.}~\bibnamefont {Sagnes}}, \emph {et~al.},\ }\href@noop {} {\bibfield  {journal} {\bibinfo  {journal} {Nature Photonics}\ }\textbf {\bibinfo {volume} {18}},\ \bibinfo {pages} {603} (\bibinfo {year} {2024})}\BibitemShut {NoStop}%
\bibitem [{\citenamefont {Flamini}\ \emph {et~al.}(2018)\citenamefont {Flamini}, \citenamefont {Spagnolo},\ and\ \citenamefont {Sciarrino}}]{flamini2018photonic}%
  \BibitemOpen
  \bibfield  {author} {\bibinfo {author} {\bibfnamefont {F.}~\bibnamefont {Flamini}}, \bibinfo {author} {\bibfnamefont {N.}~\bibnamefont {Spagnolo}},\ and\ \bibinfo {author} {\bibfnamefont {F.}~\bibnamefont {Sciarrino}},\ }\href@noop {} {\bibfield  {journal} {\bibinfo  {journal} {Reports on Progress in Physics}\ }\textbf {\bibinfo {volume} {82}},\ \bibinfo {pages} {016001} (\bibinfo {year} {2018})}\BibitemShut {NoStop}%
\bibitem [{\citenamefont {Hadfield}(2009)}]{hadfield2009single}%
  \BibitemOpen
  \bibfield  {author} {\bibinfo {author} {\bibfnamefont {R.~H.}\ \bibnamefont {Hadfield}},\ }\href@noop {} {\bibfield  {journal} {\bibinfo  {journal} {Nature photonics}\ }\textbf {\bibinfo {volume} {3}},\ \bibinfo {pages} {696} (\bibinfo {year} {2009})}\BibitemShut {NoStop}%
\bibitem [{\citenamefont {Esmaeil~Zadeh}\ \emph {et~al.}(2021)\citenamefont {Esmaeil~Zadeh}, \citenamefont {Chang}, \citenamefont {Los}, \citenamefont {Gyger}, \citenamefont {Elshaari}, \citenamefont {Steinhauer}, \citenamefont {Dorenbos},\ and\ \citenamefont {Zwiller}}]{esmaeil2021superconducting}%
  \BibitemOpen
  \bibfield  {author} {\bibinfo {author} {\bibfnamefont {I.}~\bibnamefont {Esmaeil~Zadeh}}, \bibinfo {author} {\bibfnamefont {J.}~\bibnamefont {Chang}}, \bibinfo {author} {\bibfnamefont {J.~W.}\ \bibnamefont {Los}}, \bibinfo {author} {\bibfnamefont {S.}~\bibnamefont {Gyger}}, \bibinfo {author} {\bibfnamefont {A.~W.}\ \bibnamefont {Elshaari}}, \bibinfo {author} {\bibfnamefont {S.}~\bibnamefont {Steinhauer}}, \bibinfo {author} {\bibfnamefont {S.~N.}\ \bibnamefont {Dorenbos}},\ and\ \bibinfo {author} {\bibfnamefont {V.}~\bibnamefont {Zwiller}},\ }\href@noop {} {\bibfield  {journal} {\bibinfo  {journal} {Applied Physics Letters}\ }\textbf {\bibinfo {volume} {118}} (\bibinfo {year} {2021})}\BibitemShut {NoStop}%
\bibitem [{\citenamefont {Cahall}\ \emph {et~al.}(2017)\citenamefont {Cahall}, \citenamefont {Nicolich}, \citenamefont {Islam}, \citenamefont {Lafyatis}, \citenamefont {Miller}, \citenamefont {Gauthier},\ and\ \citenamefont {Kim}}]{cahall2017multi}%
  \BibitemOpen
  \bibfield  {author} {\bibinfo {author} {\bibfnamefont {C.}~\bibnamefont {Cahall}}, \bibinfo {author} {\bibfnamefont {K.~L.}\ \bibnamefont {Nicolich}}, \bibinfo {author} {\bibfnamefont {N.~T.}\ \bibnamefont {Islam}}, \bibinfo {author} {\bibfnamefont {G.~P.}\ \bibnamefont {Lafyatis}}, \bibinfo {author} {\bibfnamefont {A.~J.}\ \bibnamefont {Miller}}, \bibinfo {author} {\bibfnamefont {D.~J.}\ \bibnamefont {Gauthier}},\ and\ \bibinfo {author} {\bibfnamefont {J.}~\bibnamefont {Kim}},\ }\href@noop {} {\bibfield  {journal} {\bibinfo  {journal} {Optica}\ }\textbf {\bibinfo {volume} {4}},\ \bibinfo {pages} {1534} (\bibinfo {year} {2017})}\BibitemShut {NoStop}%
\bibitem [{\citenamefont {Reithmaier}\ \emph {et~al.}(2013)\citenamefont {Reithmaier}, \citenamefont {Lichtmannecker}, \citenamefont {Reichert}, \citenamefont {Hasch}, \citenamefont {M{\"u}ller}, \citenamefont {Bichler}, \citenamefont {Gross},\ and\ \citenamefont {Finley}}]{reithmaier2013chip}%
  \BibitemOpen
  \bibfield  {author} {\bibinfo {author} {\bibfnamefont {G.}~\bibnamefont {Reithmaier}}, \bibinfo {author} {\bibfnamefont {S.}~\bibnamefont {Lichtmannecker}}, \bibinfo {author} {\bibfnamefont {T.}~\bibnamefont {Reichert}}, \bibinfo {author} {\bibfnamefont {P.}~\bibnamefont {Hasch}}, \bibinfo {author} {\bibfnamefont {K.}~\bibnamefont {M{\"u}ller}}, \bibinfo {author} {\bibfnamefont {M.}~\bibnamefont {Bichler}}, \bibinfo {author} {\bibfnamefont {R.}~\bibnamefont {Gross}},\ and\ \bibinfo {author} {\bibfnamefont {J.~J.}\ \bibnamefont {Finley}},\ }\href@noop {} {\bibfield  {journal} {\bibinfo  {journal} {Scientific reports}\ }\textbf {\bibinfo {volume} {3}},\ \bibinfo {pages} {1901} (\bibinfo {year} {2013})}\BibitemShut {NoStop}%
\bibitem [{\citenamefont {Cucciniello}\ \emph {et~al.}(2022)\citenamefont {Cucciniello}, \citenamefont {Lee}, \citenamefont {Feng}, \citenamefont {Yang}, \citenamefont {Zeng}, \citenamefont {Patibandla}, \citenamefont {Zhu},\ and\ \citenamefont {Jia}}]{cucciniello2022superconducting}%
  \BibitemOpen
  \bibfield  {author} {\bibinfo {author} {\bibfnamefont {N.}~\bibnamefont {Cucciniello}}, \bibinfo {author} {\bibfnamefont {D.}~\bibnamefont {Lee}}, \bibinfo {author} {\bibfnamefont {H.~Y.}\ \bibnamefont {Feng}}, \bibinfo {author} {\bibfnamefont {Z.}~\bibnamefont {Yang}}, \bibinfo {author} {\bibfnamefont {H.}~\bibnamefont {Zeng}}, \bibinfo {author} {\bibfnamefont {N.}~\bibnamefont {Patibandla}}, \bibinfo {author} {\bibfnamefont {M.}~\bibnamefont {Zhu}},\ and\ \bibinfo {author} {\bibfnamefont {Q.}~\bibnamefont {Jia}},\ }\href@noop {} {\bibfield  {journal} {\bibinfo  {journal} {Journal of Physics: Condensed Matter}\ }\textbf {\bibinfo {volume} {34}},\ \bibinfo {pages} {374003} (\bibinfo {year} {2022})}\BibitemShut {NoStop}%
\bibitem [{\citenamefont {Miki}\ \emph {et~al.}(2009)\citenamefont {Miki}, \citenamefont {Takeda}, \citenamefont {Fujiwara}, \citenamefont {Sasaki}, \citenamefont {Otomo},\ and\ \citenamefont {Wang}}]{miki2009superconducting}%
  \BibitemOpen
  \bibfield  {author} {\bibinfo {author} {\bibfnamefont {S.}~\bibnamefont {Miki}}, \bibinfo {author} {\bibfnamefont {M.}~\bibnamefont {Takeda}}, \bibinfo {author} {\bibfnamefont {M.}~\bibnamefont {Fujiwara}}, \bibinfo {author} {\bibfnamefont {M.}~\bibnamefont {Sasaki}}, \bibinfo {author} {\bibfnamefont {A.}~\bibnamefont {Otomo}},\ and\ \bibinfo {author} {\bibfnamefont {Z.}~\bibnamefont {Wang}},\ }\href@noop {} {\bibfield  {journal} {\bibinfo  {journal} {Applied physics express}\ }\textbf {\bibinfo {volume} {2}},\ \bibinfo {pages} {075002} (\bibinfo {year} {2009})}\BibitemShut {NoStop}%
\bibitem [{\citenamefont {Ilin}\ \emph {et~al.}(2008)\citenamefont {Ilin}, \citenamefont {Schneider}, \citenamefont {Gerthsen}, \citenamefont {Engel}, \citenamefont {Bartolf}, \citenamefont {Schilling}, \citenamefont {Semenov}, \citenamefont {H{\"u}bers}, \citenamefont {Freitag},\ and\ \citenamefont {Siegel}}]{ilin2008ultra}%
  \BibitemOpen
  \bibfield  {author} {\bibinfo {author} {\bibfnamefont {K.}~\bibnamefont {Ilin}}, \bibinfo {author} {\bibfnamefont {R.}~\bibnamefont {Schneider}}, \bibinfo {author} {\bibfnamefont {D.}~\bibnamefont {Gerthsen}}, \bibinfo {author} {\bibfnamefont {A.}~\bibnamefont {Engel}}, \bibinfo {author} {\bibfnamefont {H.}~\bibnamefont {Bartolf}}, \bibinfo {author} {\bibfnamefont {A.}~\bibnamefont {Schilling}}, \bibinfo {author} {\bibfnamefont {A.}~\bibnamefont {Semenov}}, \bibinfo {author} {\bibfnamefont {H.-W.}\ \bibnamefont {H{\"u}bers}}, \bibinfo {author} {\bibfnamefont {B.}~\bibnamefont {Freitag}},\ and\ \bibinfo {author} {\bibfnamefont {M.}~\bibnamefont {Siegel}},\ }in\ \href@noop {} {\emph {\bibinfo {booktitle} {Journal of Physics: Conference Series}}},\ Vol.~\bibinfo {volume} {97}\ (\bibinfo {organization} {IOP Publishing},\ \bibinfo {year} {2008})\ p.\ \bibinfo {pages} {012045}\BibitemShut {NoStop}%
\bibitem [{\citenamefont {Licata}\ \emph {et~al.}(2022)\citenamefont {Licata}, \citenamefont {Sarker}, \citenamefont {Bachhav}, \citenamefont {Roy}, \citenamefont {Wei}, \citenamefont {Yang}, \citenamefont {Patibandla}, \citenamefont {Zeng}, \citenamefont {Zhu}, \citenamefont {Jia} \emph {et~al.}}]{licata2022correlation}%
  \BibitemOpen
  \bibfield  {author} {\bibinfo {author} {\bibfnamefont {O.~G.}\ \bibnamefont {Licata}}, \bibinfo {author} {\bibfnamefont {J.}~\bibnamefont {Sarker}}, \bibinfo {author} {\bibfnamefont {M.}~\bibnamefont {Bachhav}}, \bibinfo {author} {\bibfnamefont {P.}~\bibnamefont {Roy}}, \bibinfo {author} {\bibfnamefont {X.}~\bibnamefont {Wei}}, \bibinfo {author} {\bibfnamefont {Z.}~\bibnamefont {Yang}}, \bibinfo {author} {\bibfnamefont {N.}~\bibnamefont {Patibandla}}, \bibinfo {author} {\bibfnamefont {H.}~\bibnamefont {Zeng}}, \bibinfo {author} {\bibfnamefont {M.}~\bibnamefont {Zhu}}, \bibinfo {author} {\bibfnamefont {Q.}~\bibnamefont {Jia}}, \emph {et~al.},\ }\href@noop {} {\bibfield  {journal} {\bibinfo  {journal} {Materials Chemistry and Physics}\ }\textbf {\bibinfo {volume} {282}},\ \bibinfo {pages} {125962} (\bibinfo {year} {2022})}\BibitemShut {NoStop}%
\bibitem [{\citenamefont {Wang}\ \emph {et~al.}(2022)\citenamefont {Wang}, \citenamefont {Guo}, \citenamefont {Miao}, \citenamefont {Luo}, \citenamefont {Gu}, \citenamefont {Xie}, \citenamefont {Wang}, \citenamefont {Zhang}, \citenamefont {Wang},\ and\ \citenamefont {Hu}}]{wang2022emerging}%
  \BibitemOpen
  \bibfield  {author} {\bibinfo {author} {\bibfnamefont {H.}~\bibnamefont {Wang}}, \bibinfo {author} {\bibfnamefont {J.}~\bibnamefont {Guo}}, \bibinfo {author} {\bibfnamefont {J.}~\bibnamefont {Miao}}, \bibinfo {author} {\bibfnamefont {W.}~\bibnamefont {Luo}}, \bibinfo {author} {\bibfnamefont {Y.}~\bibnamefont {Gu}}, \bibinfo {author} {\bibfnamefont {R.}~\bibnamefont {Xie}}, \bibinfo {author} {\bibfnamefont {F.}~\bibnamefont {Wang}}, \bibinfo {author} {\bibfnamefont {L.}~\bibnamefont {Zhang}}, \bibinfo {author} {\bibfnamefont {P.}~\bibnamefont {Wang}},\ and\ \bibinfo {author} {\bibfnamefont {W.}~\bibnamefont {Hu}},\ }\href@noop {} {\bibfield  {journal} {\bibinfo  {journal} {Small}\ }\textbf {\bibinfo {volume} {18}},\ \bibinfo {pages} {2103963} (\bibinfo {year} {2022})}\BibitemShut {NoStop}%
\bibitem [{\citenamefont {Di~Battista}\ \emph {et~al.}(2024)\citenamefont {Di~Battista}, \citenamefont {Fong}, \citenamefont {D{\'\i}ez-Carl{\'o}n}, \citenamefont {Watanabe}, \citenamefont {Taniguchi},\ and\ \citenamefont {Efetov}}]{di2024infrared}%
  \BibitemOpen
  \bibfield  {author} {\bibinfo {author} {\bibfnamefont {G.}~\bibnamefont {Di~Battista}}, \bibinfo {author} {\bibfnamefont {K.~C.}\ \bibnamefont {Fong}}, \bibinfo {author} {\bibfnamefont {A.}~\bibnamefont {D{\'\i}ez-Carl{\'o}n}}, \bibinfo {author} {\bibfnamefont {K.}~\bibnamefont {Watanabe}}, \bibinfo {author} {\bibfnamefont {T.}~\bibnamefont {Taniguchi}},\ and\ \bibinfo {author} {\bibfnamefont {D.~K.}\ \bibnamefont {Efetov}},\ }\href@noop {} {\bibfield  {journal} {\bibinfo  {journal} {Science Advances}\ }\textbf {\bibinfo {volume} {10}},\ \bibinfo {pages} {eadp3725} (\bibinfo {year} {2024})}\BibitemShut {NoStop}%
\bibitem [{\citenamefont {Montblanch}\ \emph {et~al.}(2023)\citenamefont {Montblanch}, \citenamefont {Barbone}, \citenamefont {Aharonovich}, \citenamefont {Atat{\"u}re},\ and\ \citenamefont {Ferrari}}]{montblanch2023layered}%
  \BibitemOpen
  \bibfield  {author} {\bibinfo {author} {\bibfnamefont {A.~R.-P.}\ \bibnamefont {Montblanch}}, \bibinfo {author} {\bibfnamefont {M.}~\bibnamefont {Barbone}}, \bibinfo {author} {\bibfnamefont {I.}~\bibnamefont {Aharonovich}}, \bibinfo {author} {\bibfnamefont {M.}~\bibnamefont {Atat{\"u}re}},\ and\ \bibinfo {author} {\bibfnamefont {A.~C.}\ \bibnamefont {Ferrari}},\ }\href@noop {} {\bibfield  {journal} {\bibinfo  {journal} {Nature Nanotechnology}\ }\textbf {\bibinfo {volume} {18}},\ \bibinfo {pages} {555} (\bibinfo {year} {2023})}\BibitemShut {NoStop}%
\bibitem [{\citenamefont {Staley}\ \emph {et~al.}(2009)\citenamefont {Staley}, \citenamefont {Wu}, \citenamefont {Eklund}, \citenamefont {Liu}, \citenamefont {Li},\ and\ \citenamefont {Xu}}]{staley2009electric}%
  \BibitemOpen
  \bibfield  {author} {\bibinfo {author} {\bibfnamefont {N.~E.}\ \bibnamefont {Staley}}, \bibinfo {author} {\bibfnamefont {J.}~\bibnamefont {Wu}}, \bibinfo {author} {\bibfnamefont {P.}~\bibnamefont {Eklund}}, \bibinfo {author} {\bibfnamefont {Y.}~\bibnamefont {Liu}}, \bibinfo {author} {\bibfnamefont {L.}~\bibnamefont {Li}},\ and\ \bibinfo {author} {\bibfnamefont {Z.}~\bibnamefont {Xu}},\ }\href@noop {} {\bibfield  {journal} {\bibinfo  {journal} {Physical Review B—Condensed Matter and Materials Physics}\ }\textbf {\bibinfo {volume} {80}},\ \bibinfo {pages} {184505} (\bibinfo {year} {2009})}\BibitemShut {NoStop}%
\bibitem [{\citenamefont {Mills}\ \emph {et~al.}(2014)\citenamefont {Mills}, \citenamefont {Staley}, \citenamefont {Wisser}, \citenamefont {Shen}, \citenamefont {Xu},\ and\ \citenamefont {Liu}}]{mills2014single}%
  \BibitemOpen
  \bibfield  {author} {\bibinfo {author} {\bibfnamefont {S.~A.}\ \bibnamefont {Mills}}, \bibinfo {author} {\bibfnamefont {N.~E.}\ \bibnamefont {Staley}}, \bibinfo {author} {\bibfnamefont {J.~J.}\ \bibnamefont {Wisser}}, \bibinfo {author} {\bibfnamefont {C.}~\bibnamefont {Shen}}, \bibinfo {author} {\bibfnamefont {Z.}~\bibnamefont {Xu}},\ and\ \bibinfo {author} {\bibfnamefont {Y.}~\bibnamefont {Liu}},\ }\href@noop {} {\bibfield  {journal} {\bibinfo  {journal} {Applied Physics Letters}\ }\textbf {\bibinfo {volume} {104}} (\bibinfo {year} {2014})}\BibitemShut {NoStop}%
\bibitem [{\citenamefont {Paralikis}\ \emph {et~al.}(2024)\citenamefont {Paralikis}, \citenamefont {Piccinini}, \citenamefont {Madigawa}, \citenamefont {Metuh}, \citenamefont {Vannucci}, \citenamefont {Gregersen},\ and\ \citenamefont {Munkhbat}}]{paralikis2024tailoring}%
  \BibitemOpen
  \bibfield  {author} {\bibinfo {author} {\bibfnamefont {A.}~\bibnamefont {Paralikis}}, \bibinfo {author} {\bibfnamefont {C.}~\bibnamefont {Piccinini}}, \bibinfo {author} {\bibfnamefont {A.~A.}\ \bibnamefont {Madigawa}}, \bibinfo {author} {\bibfnamefont {P.}~\bibnamefont {Metuh}}, \bibinfo {author} {\bibfnamefont {L.}~\bibnamefont {Vannucci}}, \bibinfo {author} {\bibfnamefont {N.}~\bibnamefont {Gregersen}},\ and\ \bibinfo {author} {\bibfnamefont {B.}~\bibnamefont {Munkhbat}},\ }\href@noop {} {\bibfield  {journal} {\bibinfo  {journal} {npj 2D Materials and Applications}\ }\textbf {\bibinfo {volume} {8}},\ \bibinfo {pages} {59} (\bibinfo {year} {2024})}\BibitemShut {NoStop}%
\bibitem [{\citenamefont {Munkhbat}\ \emph {et~al.}(2023)\citenamefont {Munkhbat}, \citenamefont {K{\"u}{\c{c}}{\"u}k{\"o}z}, \citenamefont {Baranov}, \citenamefont {Antosiewicz},\ and\ \citenamefont {Shegai}}]{munkhbat2023nanostructured}%
  \BibitemOpen
  \bibfield  {author} {\bibinfo {author} {\bibfnamefont {B.}~\bibnamefont {Munkhbat}}, \bibinfo {author} {\bibfnamefont {B.}~\bibnamefont {K{\"u}{\c{c}}{\"u}k{\"o}z}}, \bibinfo {author} {\bibfnamefont {D.~G.}\ \bibnamefont {Baranov}}, \bibinfo {author} {\bibfnamefont {T.~J.}\ \bibnamefont {Antosiewicz}},\ and\ \bibinfo {author} {\bibfnamefont {T.~O.}\ \bibnamefont {Shegai}},\ }\href@noop {} {\bibfield  {journal} {\bibinfo  {journal} {Laser \& Photonics Reviews}\ }\textbf {\bibinfo {volume} {17}},\ \bibinfo {pages} {2200057} (\bibinfo {year} {2023})}\BibitemShut {NoStop}%
\bibitem [{\citenamefont {Frindt}(1972)}]{frindt1972superconductivity}%
  \BibitemOpen
  \bibfield  {author} {\bibinfo {author} {\bibfnamefont {R.}~\bibnamefont {Frindt}},\ }\href@noop {} {\bibfield  {journal} {\bibinfo  {journal} {Physical Review Letters}\ }\textbf {\bibinfo {volume} {28}},\ \bibinfo {pages} {299} (\bibinfo {year} {1972})}\BibitemShut {NoStop}%
\bibitem [{\citenamefont {Xi}\ \emph {et~al.}(2016)\citenamefont {Xi}, \citenamefont {Wang}, \citenamefont {Zhao}, \citenamefont {Park}, \citenamefont {Law}, \citenamefont {Berger}, \citenamefont {Forr{\'o}}, \citenamefont {Shan},\ and\ \citenamefont {Mak}}]{xi2016ising}%
  \BibitemOpen
  \bibfield  {author} {\bibinfo {author} {\bibfnamefont {X.}~\bibnamefont {Xi}}, \bibinfo {author} {\bibfnamefont {Z.}~\bibnamefont {Wang}}, \bibinfo {author} {\bibfnamefont {W.}~\bibnamefont {Zhao}}, \bibinfo {author} {\bibfnamefont {J.-H.}\ \bibnamefont {Park}}, \bibinfo {author} {\bibfnamefont {K.~T.}\ \bibnamefont {Law}}, \bibinfo {author} {\bibfnamefont {H.}~\bibnamefont {Berger}}, \bibinfo {author} {\bibfnamefont {L.}~\bibnamefont {Forr{\'o}}}, \bibinfo {author} {\bibfnamefont {J.}~\bibnamefont {Shan}},\ and\ \bibinfo {author} {\bibfnamefont {K.~F.}\ \bibnamefont {Mak}},\ }\href@noop {} {\bibfield  {journal} {\bibinfo  {journal} {Nature Physics}\ }\textbf {\bibinfo {volume} {12}},\ \bibinfo {pages} {139} (\bibinfo {year} {2016})}\BibitemShut {NoStop}%
\bibitem [{\citenamefont {Tian}\ \emph {et~al.}(2020)\citenamefont {Tian}, \citenamefont {Bottala-Gambetta}, \citenamefont {Marchetto}, \citenamefont {Jacquemin}, \citenamefont {Crisci}, \citenamefont {Reboud}, \citenamefont {Mantoux}, \citenamefont {Berthom{\'e}}, \citenamefont {Mercier}, \citenamefont {Sulpice} \emph {et~al.}}]{tian2020improved}%
  \BibitemOpen
  \bibfield  {author} {\bibinfo {author} {\bibfnamefont {L.}~\bibnamefont {Tian}}, \bibinfo {author} {\bibfnamefont {I.}~\bibnamefont {Bottala-Gambetta}}, \bibinfo {author} {\bibfnamefont {V.}~\bibnamefont {Marchetto}}, \bibinfo {author} {\bibfnamefont {M.}~\bibnamefont {Jacquemin}}, \bibinfo {author} {\bibfnamefont {A.}~\bibnamefont {Crisci}}, \bibinfo {author} {\bibfnamefont {R.}~\bibnamefont {Reboud}}, \bibinfo {author} {\bibfnamefont {A.}~\bibnamefont {Mantoux}}, \bibinfo {author} {\bibfnamefont {G.}~\bibnamefont {Berthom{\'e}}}, \bibinfo {author} {\bibfnamefont {F.}~\bibnamefont {Mercier}}, \bibinfo {author} {\bibfnamefont {A.}~\bibnamefont {Sulpice}}, \emph {et~al.},\ }\href@noop {} {\bibfield  {journal} {\bibinfo  {journal} {Thin Solid Films}\ }\textbf {\bibinfo {volume} {709}},\ \bibinfo {pages} {138232} (\bibinfo {year} {2020})}\BibitemShut {NoStop}%
\bibitem [{\citenamefont {Cukauskas}(1983)}]{cukauskas1983effects}%
  \BibitemOpen
  \bibfield  {author} {\bibinfo {author} {\bibfnamefont {E.}~\bibnamefont {Cukauskas}},\ }\href@noop {} {\bibfield  {journal} {\bibinfo  {journal} {Journal of Applied Physics}\ }\textbf {\bibinfo {volume} {54}},\ \bibinfo {pages} {1013} (\bibinfo {year} {1983})}\BibitemShut {NoStop}%
\bibitem [{\citenamefont {Shein}\ \emph {et~al.}(2024)\citenamefont {Shein}, \citenamefont {Zharkova}, \citenamefont {Kashchenko}, \citenamefont {Kolbatova}, \citenamefont {Lyubchak}, \citenamefont {Elesin}, \citenamefont {Nguyen}, \citenamefont {Semenov}, \citenamefont {Charaev}, \citenamefont {Schilling} \emph {et~al.}}]{shein2024fundamental}%
  \BibitemOpen
  \bibfield  {author} {\bibinfo {author} {\bibfnamefont {K.}~\bibnamefont {Shein}}, \bibinfo {author} {\bibfnamefont {E.}~\bibnamefont {Zharkova}}, \bibinfo {author} {\bibfnamefont {M.}~\bibnamefont {Kashchenko}}, \bibinfo {author} {\bibfnamefont {A.}~\bibnamefont {Kolbatova}}, \bibinfo {author} {\bibfnamefont {A.}~\bibnamefont {Lyubchak}}, \bibinfo {author} {\bibfnamefont {L.}~\bibnamefont {Elesin}}, \bibinfo {author} {\bibfnamefont {E.}~\bibnamefont {Nguyen}}, \bibinfo {author} {\bibfnamefont {A.}~\bibnamefont {Semenov}}, \bibinfo {author} {\bibfnamefont {I.}~\bibnamefont {Charaev}}, \bibinfo {author} {\bibfnamefont {A.}~\bibnamefont {Schilling}}, \emph {et~al.},\ }\href@noop {} {\bibfield  {journal} {\bibinfo  {journal} {Nano letters}\ }\textbf {\bibinfo {volume} {24}},\ \bibinfo {pages} {2282} (\bibinfo {year} {2024})}\BibitemShut {NoStop}%
\bibitem [{\citenamefont {Orchin}(2021)}]{orchin2021two}%
  \BibitemOpen
  \bibfield  {author} {\bibinfo {author} {\bibfnamefont {G.~J.}\ \bibnamefont {Orchin}},\ }\emph {\bibinfo {title} {Two dimensional superconductors for infrared photodetection}},\ \href@noop {} {Ph.D. thesis},\ \bibinfo  {school} {University of Glasgow} (\bibinfo {year} {2021})\BibitemShut {NoStop}%
\bibitem [{\citenamefont {El-Bana}\ \emph {et~al.}(2013)\citenamefont {El-Bana}, \citenamefont {Wolverson}, \citenamefont {Russo}, \citenamefont {Balakrishnan}, \citenamefont {Paul},\ and\ \citenamefont {Bending}}]{el2013superconductivity}%
  \BibitemOpen
  \bibfield  {author} {\bibinfo {author} {\bibfnamefont {M.~S.}\ \bibnamefont {El-Bana}}, \bibinfo {author} {\bibfnamefont {D.}~\bibnamefont {Wolverson}}, \bibinfo {author} {\bibfnamefont {S.}~\bibnamefont {Russo}}, \bibinfo {author} {\bibfnamefont {G.}~\bibnamefont {Balakrishnan}}, \bibinfo {author} {\bibfnamefont {D.~M.}\ \bibnamefont {Paul}},\ and\ \bibinfo {author} {\bibfnamefont {S.~J.}\ \bibnamefont {Bending}},\ }\href@noop {} {\bibfield  {journal} {\bibinfo  {journal} {Superconductor Science and Technology}\ }\textbf {\bibinfo {volume} {26}},\ \bibinfo {pages} {125020} (\bibinfo {year} {2013})}\BibitemShut {NoStop}%
\bibitem [{\citenamefont {Orchin}\ \emph {et~al.}(2019)\citenamefont {Orchin}, \citenamefont {De~Fazio}, \citenamefont {Di~Bernardo}, \citenamefont {Hamer}, \citenamefont {Yoon}, \citenamefont {Cadore}, \citenamefont {Goykhman}, \citenamefont {Watanabe}, \citenamefont {Taniguchi}, \citenamefont {Robinson} \emph {et~al.}}]{orchin2019niobium}%
  \BibitemOpen
  \bibfield  {author} {\bibinfo {author} {\bibfnamefont {G.~J.}\ \bibnamefont {Orchin}}, \bibinfo {author} {\bibfnamefont {D.}~\bibnamefont {De~Fazio}}, \bibinfo {author} {\bibfnamefont {A.}~\bibnamefont {Di~Bernardo}}, \bibinfo {author} {\bibfnamefont {M.}~\bibnamefont {Hamer}}, \bibinfo {author} {\bibfnamefont {D.}~\bibnamefont {Yoon}}, \bibinfo {author} {\bibfnamefont {A.~R.}\ \bibnamefont {Cadore}}, \bibinfo {author} {\bibfnamefont {I.}~\bibnamefont {Goykhman}}, \bibinfo {author} {\bibfnamefont {K.}~\bibnamefont {Watanabe}}, \bibinfo {author} {\bibfnamefont {T.}~\bibnamefont {Taniguchi}}, \bibinfo {author} {\bibfnamefont {J.~W.}\ \bibnamefont {Robinson}}, \emph {et~al.},\ }\href@noop {} {\bibfield  {journal} {\bibinfo  {journal} {Applied Physics Letters}\ }\textbf {\bibinfo {volume} {114}} (\bibinfo {year} {2019})}\BibitemShut {NoStop}%
\bibitem [{\citenamefont {Shein~Kirill}\ \emph {et~al.}(2024)\citenamefont {Shein~Kirill}, \citenamefont {Lyubchak~Anastasia}, \citenamefont {Ekaterina}, \citenamefont {Bandurin~Denis}, \citenamefont {Ilya}, \citenamefont {Gayduchenko~Igor},\ and\ \citenamefont {Goltsman~Grigory}}]{shein2024towards}%
  \BibitemOpen
  \bibfield  {author} {\bibinfo {author} {\bibfnamefont {V.}~\bibnamefont {Shein~Kirill}}, \bibinfo {author} {\bibfnamefont {N.}~\bibnamefont {Lyubchak~Anastasia}}, \bibinfo {author} {\bibfnamefont {Z.}~\bibnamefont {Ekaterina}}, \bibinfo {author} {\bibfnamefont {A.}~\bibnamefont {Bandurin~Denis}}, \bibinfo {author} {\bibfnamefont {C.}~\bibnamefont {Ilya}}, \bibinfo {author} {\bibfnamefont {A.}~\bibnamefont {Gayduchenko~Igor}},\ and\ \bibinfo {author} {\bibfnamefont {N.}~\bibnamefont {Goltsman~Grigory}},\ }\href@noop {} {\bibfield  {journal} {\bibinfo  {journal} {St. Petersburg Polytechnic University Journal: Physics and Mathematics}\ }\textbf {\bibinfo {volume} {76}},\ \bibinfo {pages} {241} (\bibinfo {year} {2024})}\BibitemShut {NoStop}%
\bibitem [{\citenamefont {Jayanand}\ \emph {et~al.}(2024)\citenamefont {Jayanand}, \citenamefont {Saenz}, \citenamefont {Krylyuk}, \citenamefont {Davydov}, \citenamefont {Karapetrov}, \citenamefont {Liu}, \citenamefont {Zhou},\ and\ \citenamefont {Kaul}}]{jayanand2024optically}%
  \BibitemOpen
  \bibfield  {author} {\bibinfo {author} {\bibfnamefont {K.}~\bibnamefont {Jayanand}}, \bibinfo {author} {\bibfnamefont {G.~A.}\ \bibnamefont {Saenz}}, \bibinfo {author} {\bibfnamefont {S.}~\bibnamefont {Krylyuk}}, \bibinfo {author} {\bibfnamefont {A.~V.}\ \bibnamefont {Davydov}}, \bibinfo {author} {\bibfnamefont {G.}~\bibnamefont {Karapetrov}}, \bibinfo {author} {\bibfnamefont {Z.}~\bibnamefont {Liu}}, \bibinfo {author} {\bibfnamefont {W.}~\bibnamefont {Zhou}},\ and\ \bibinfo {author} {\bibfnamefont {A.~B.}\ \bibnamefont {Kaul}},\ }\href@noop {} {\bibfield  {journal} {\bibinfo  {journal} {Iscience}\ }\textbf {\bibinfo {volume} {27}} (\bibinfo {year} {2024})}\BibitemShut {NoStop}%
\bibitem [{\citenamefont {Jiang}\ \emph {et~al.}(2023)\citenamefont {Jiang}, \citenamefont {Xing}, \citenamefont {Zhang}, \citenamefont {Han}, \citenamefont {Zhang}, \citenamefont {Yao}, \citenamefont {Wang}, \citenamefont {Wang}, \citenamefont {Lan}, \citenamefont {Lv} \emph {et~al.}}]{jiang2023fast}%
  \BibitemOpen
  \bibfield  {author} {\bibinfo {author} {\bibfnamefont {M.}~\bibnamefont {Jiang}}, \bibinfo {author} {\bibfnamefont {H.}~\bibnamefont {Xing}}, \bibinfo {author} {\bibfnamefont {L.}~\bibnamefont {Zhang}}, \bibinfo {author} {\bibfnamefont {L.}~\bibnamefont {Han}}, \bibinfo {author} {\bibfnamefont {K.}~\bibnamefont {Zhang}}, \bibinfo {author} {\bibfnamefont {C.}~\bibnamefont {Yao}}, \bibinfo {author} {\bibfnamefont {D.}~\bibnamefont {Wang}}, \bibinfo {author} {\bibfnamefont {X.}~\bibnamefont {Wang}}, \bibinfo {author} {\bibfnamefont {S.}~\bibnamefont {Lan}}, \bibinfo {author} {\bibfnamefont {X.}~\bibnamefont {Lv}}, \emph {et~al.},\ }\href@noop {} {\bibfield  {journal} {\bibinfo  {journal} {Advanced Optical Materials}\ }\textbf {\bibinfo {volume} {11}},\ \bibinfo {pages} {2300074} (\bibinfo {year} {2023})}\BibitemShut {NoStop}%
\bibitem [{\citenamefont {Tinkham}(1996)}]{Tinkham1996}%
  \BibitemOpen
  \bibfield  {author} {\bibinfo {author} {\bibfnamefont {M.}~\bibnamefont {Tinkham}},\ }\href@noop {} {\emph {\bibinfo {title} {Introduction to Superconductivity}}}\ (\bibinfo  {publisher} {Dover Publications},\ \bibinfo {year} {1996})\BibitemShut {NoStop}%
\bibitem [{\citenamefont {Jo}\ \emph {et~al.}(2013)\citenamefont {Jo}, \citenamefont {Pettes}, \citenamefont {Kim}, \citenamefont {Watanabe}, \citenamefont {Taniguchi}, \citenamefont {Yao},\ and\ \citenamefont {Shi}}]{jo2013thermal}%
  \BibitemOpen
  \bibfield  {author} {\bibinfo {author} {\bibfnamefont {I.}~\bibnamefont {Jo}}, \bibinfo {author} {\bibfnamefont {M.~T.}\ \bibnamefont {Pettes}}, \bibinfo {author} {\bibfnamefont {J.}~\bibnamefont {Kim}}, \bibinfo {author} {\bibfnamefont {K.}~\bibnamefont {Watanabe}}, \bibinfo {author} {\bibfnamefont {T.}~\bibnamefont {Taniguchi}}, \bibinfo {author} {\bibfnamefont {Z.}~\bibnamefont {Yao}},\ and\ \bibinfo {author} {\bibfnamefont {L.}~\bibnamefont {Shi}},\ }\href@noop {} {\bibfield  {journal} {\bibinfo  {journal} {Nano letters}\ }\textbf {\bibinfo {volume} {13}},\ \bibinfo {pages} {550} (\bibinfo {year} {2013})}\BibitemShut {NoStop}%
\bibitem [{\citenamefont {Sichel}\ \emph {et~al.}(1976)\citenamefont {Sichel}, \citenamefont {Miller}, \citenamefont {Abrahams},\ and\ \citenamefont {Buiocchi}}]{sichel1976heat}%
  \BibitemOpen
  \bibfield  {author} {\bibinfo {author} {\bibfnamefont {E.}~\bibnamefont {Sichel}}, \bibinfo {author} {\bibfnamefont {R.}~\bibnamefont {Miller}}, \bibinfo {author} {\bibfnamefont {M.}~\bibnamefont {Abrahams}},\ and\ \bibinfo {author} {\bibfnamefont {C.}~\bibnamefont {Buiocchi}},\ }\href@noop {} {\bibfield  {journal} {\bibinfo  {journal} {Physical review B}\ }\textbf {\bibinfo {volume} {13}},\ \bibinfo {pages} {4607} (\bibinfo {year} {1976})}\BibitemShut {NoStop}%
\bibitem [{\citenamefont {Li}\ \emph {et~al.}(2011)\citenamefont {Li}, \citenamefont {Wu}, \citenamefont {Bomze}, \citenamefont {Borzenets}, \citenamefont {Finkelstein},\ and\ \citenamefont {Chang}}]{li2011retrapping}%
  \BibitemOpen
  \bibfield  {author} {\bibinfo {author} {\bibfnamefont {P.}~\bibnamefont {Li}}, \bibinfo {author} {\bibfnamefont {P.~M.}\ \bibnamefont {Wu}}, \bibinfo {author} {\bibfnamefont {Y.}~\bibnamefont {Bomze}}, \bibinfo {author} {\bibfnamefont {I.~V.}\ \bibnamefont {Borzenets}}, \bibinfo {author} {\bibfnamefont {G.}~\bibnamefont {Finkelstein}},\ and\ \bibinfo {author} {\bibfnamefont {A.}~\bibnamefont {Chang}},\ }\href@noop {} {\bibfield  {journal} {\bibinfo  {journal} {Physical Review B—Condensed Matter and Materials Physics}\ }\textbf {\bibinfo {volume} {84}},\ \bibinfo {pages} {184508} (\bibinfo {year} {2011})}\BibitemShut {NoStop}%
\bibitem [{\citenamefont {Sahu}\ \emph {et~al.}(2009)\citenamefont {Sahu}, \citenamefont {Bae}, \citenamefont {Rogachev}, \citenamefont {Pekker}, \citenamefont {Wei}, \citenamefont {Shah}, \citenamefont {Goldbart},\ and\ \citenamefont {Bezryadin}}]{sahu2009individual}%
  \BibitemOpen
  \bibfield  {author} {\bibinfo {author} {\bibfnamefont {M.}~\bibnamefont {Sahu}}, \bibinfo {author} {\bibfnamefont {M.-H.}\ \bibnamefont {Bae}}, \bibinfo {author} {\bibfnamefont {A.}~\bibnamefont {Rogachev}}, \bibinfo {author} {\bibfnamefont {D.}~\bibnamefont {Pekker}}, \bibinfo {author} {\bibfnamefont {T.-C.}\ \bibnamefont {Wei}}, \bibinfo {author} {\bibfnamefont {N.}~\bibnamefont {Shah}}, \bibinfo {author} {\bibfnamefont {P.~M.}\ \bibnamefont {Goldbart}},\ and\ \bibinfo {author} {\bibfnamefont {A.}~\bibnamefont {Bezryadin}},\ }\href@noop {} {\bibfield  {journal} {\bibinfo  {journal} {Nature Physics}\ }\textbf {\bibinfo {volume} {5}},\ \bibinfo {pages} {503} (\bibinfo {year} {2009})}\BibitemShut {NoStop}%
\bibitem [{\citenamefont {Natarajan}\ \emph {et~al.}(2012)\citenamefont {Natarajan}, \citenamefont {Tanner},\ and\ \citenamefont {Hadfield}}]{natarajan2012superconducting}%
  \BibitemOpen
  \bibfield  {author} {\bibinfo {author} {\bibfnamefont {C.~M.}\ \bibnamefont {Natarajan}}, \bibinfo {author} {\bibfnamefont {M.~G.}\ \bibnamefont {Tanner}},\ and\ \bibinfo {author} {\bibfnamefont {R.~H.}\ \bibnamefont {Hadfield}},\ }\href@noop {} {\bibfield  {journal} {\bibinfo  {journal} {Superconductor science and technology}\ }\textbf {\bibinfo {volume} {25}},\ \bibinfo {pages} {063001} (\bibinfo {year} {2012})}\BibitemShut {NoStop}%
\bibitem [{\citenamefont {Annunziata}\ \emph {et~al.}(2010)\citenamefont {Annunziata}, \citenamefont {Quaranta}, \citenamefont {Santavicca}, \citenamefont {Casaburi}, \citenamefont {Frunzio}, \citenamefont {Ejrnaes}, \citenamefont {Rooks}, \citenamefont {Cristiano}, \citenamefont {Pagano}, \citenamefont {Frydman} \emph {et~al.}}]{annunziata2010reset}%
  \BibitemOpen
  \bibfield  {author} {\bibinfo {author} {\bibfnamefont {A.~J.}\ \bibnamefont {Annunziata}}, \bibinfo {author} {\bibfnamefont {O.}~\bibnamefont {Quaranta}}, \bibinfo {author} {\bibfnamefont {D.~F.}\ \bibnamefont {Santavicca}}, \bibinfo {author} {\bibfnamefont {A.}~\bibnamefont {Casaburi}}, \bibinfo {author} {\bibfnamefont {L.}~\bibnamefont {Frunzio}}, \bibinfo {author} {\bibfnamefont {M.}~\bibnamefont {Ejrnaes}}, \bibinfo {author} {\bibfnamefont {M.~J.}\ \bibnamefont {Rooks}}, \bibinfo {author} {\bibfnamefont {R.}~\bibnamefont {Cristiano}}, \bibinfo {author} {\bibfnamefont {S.}~\bibnamefont {Pagano}}, \bibinfo {author} {\bibfnamefont {A.}~\bibnamefont {Frydman}}, \emph {et~al.},\ }\href@noop {} {\bibfield  {journal} {\bibinfo  {journal} {Journal of Applied Physics}\ }\textbf {\bibinfo {volume} {108}} (\bibinfo {year} {2010})}\BibitemShut {NoStop}%
\bibitem [{\citenamefont {Morozov}\ \emph {et~al.}(2021)\citenamefont {Morozov}, \citenamefont {Casaburi},\ and\ \citenamefont {Hadfield}}]{morozov2021superconducting}%
  \BibitemOpen
  \bibfield  {author} {\bibinfo {author} {\bibfnamefont {D.~V.}\ \bibnamefont {Morozov}}, \bibinfo {author} {\bibfnamefont {A.}~\bibnamefont {Casaburi}},\ and\ \bibinfo {author} {\bibfnamefont {R.~H.}\ \bibnamefont {Hadfield}},\ }\href@noop {} {\bibfield  {journal} {\bibinfo  {journal} {Contemporary Physics}\ }\textbf {\bibinfo {volume} {62}},\ \bibinfo {pages} {69} (\bibinfo {year} {2021})}\BibitemShut {NoStop}%
\bibitem [{\citenamefont {Yin}\ \emph {et~al.}(2024)\citenamefont {Yin}, \citenamefont {Wang}, \citenamefont {Wang}, \citenamefont {Yin}, \citenamefont {Chen}, \citenamefont {Jia}, \citenamefont {Wang}, \citenamefont {Zhang},\ and\ \citenamefont {Wu}}]{yin2024heat}%
  \BibitemOpen
  \bibfield  {author} {\bibinfo {author} {\bibfnamefont {W.}~\bibnamefont {Yin}}, \bibinfo {author} {\bibfnamefont {H.}~\bibnamefont {Wang}}, \bibinfo {author} {\bibfnamefont {X.}~\bibnamefont {Wang}}, \bibinfo {author} {\bibfnamefont {R.}~\bibnamefont {Yin}}, \bibinfo {author} {\bibfnamefont {Q.}~\bibnamefont {Chen}}, \bibinfo {author} {\bibfnamefont {X.}~\bibnamefont {Jia}}, \bibinfo {author} {\bibfnamefont {H.}~\bibnamefont {Wang}}, \bibinfo {author} {\bibfnamefont {L.}~\bibnamefont {Zhang}},\ and\ \bibinfo {author} {\bibfnamefont {P.}~\bibnamefont {Wu}},\ }\href@noop {} {\bibfield  {journal} {\bibinfo  {journal} {Superconductor Science and Technology}\ }\textbf {\bibinfo {volume} {37}},\ \bibinfo {pages} {073001} (\bibinfo {year} {2024})}\BibitemShut {NoStop}%
\bibitem [{\citenamefont {Gol’Tsman}\ \emph {et~al.}(2001)\citenamefont {Gol’Tsman}, \citenamefont {Okunev}, \citenamefont {Chulkova}, \citenamefont {Lipatov}, \citenamefont {Semenov}, \citenamefont {Smirnov}, \citenamefont {Voronov}, \citenamefont {Dzardanov}, \citenamefont {Williams},\ and\ \citenamefont {Sobolewski}}]{gol2001picosecond}%
  \BibitemOpen
  \bibfield  {author} {\bibinfo {author} {\bibfnamefont {G.}~\bibnamefont {Gol’Tsman}}, \bibinfo {author} {\bibfnamefont {O.}~\bibnamefont {Okunev}}, \bibinfo {author} {\bibfnamefont {G.}~\bibnamefont {Chulkova}}, \bibinfo {author} {\bibfnamefont {A.}~\bibnamefont {Lipatov}}, \bibinfo {author} {\bibfnamefont {A.}~\bibnamefont {Semenov}}, \bibinfo {author} {\bibfnamefont {K.}~\bibnamefont {Smirnov}}, \bibinfo {author} {\bibfnamefont {B.}~\bibnamefont {Voronov}}, \bibinfo {author} {\bibfnamefont {A.}~\bibnamefont {Dzardanov}}, \bibinfo {author} {\bibfnamefont {C.}~\bibnamefont {Williams}},\ and\ \bibinfo {author} {\bibfnamefont {R.}~\bibnamefont {Sobolewski}},\ }\href@noop {} {\bibfield  {journal} {\bibinfo  {journal} {Applied physics letters}\ }\textbf {\bibinfo {volume} {79}},\ \bibinfo {pages} {705} (\bibinfo {year} {2001})}\BibitemShut {NoStop}%
\bibitem [{\citenamefont {Fitzpatrick}\ \emph {et~al.}(2010)\citenamefont {Fitzpatrick}, \citenamefont {Natarajan}, \citenamefont {Warburton}, \citenamefont {Buller}, \citenamefont {Baek}, \citenamefont {Nam}, \citenamefont {Miki}, \citenamefont {Wang}, \citenamefont {Sasaki}, \citenamefont {Sinclair} \emph {et~al.}}]{fitzpatrick2010superconducting}%
  \BibitemOpen
  \bibfield  {author} {\bibinfo {author} {\bibfnamefont {C.}~\bibnamefont {Fitzpatrick}}, \bibinfo {author} {\bibfnamefont {C.}~\bibnamefont {Natarajan}}, \bibinfo {author} {\bibfnamefont {R.~E.}\ \bibnamefont {Warburton}}, \bibinfo {author} {\bibfnamefont {G.}~\bibnamefont {Buller}}, \bibinfo {author} {\bibfnamefont {B.}~\bibnamefont {Baek}}, \bibinfo {author} {\bibfnamefont {S.}~\bibnamefont {Nam}}, \bibinfo {author} {\bibfnamefont {S.}~\bibnamefont {Miki}}, \bibinfo {author} {\bibfnamefont {Z.}~\bibnamefont {Wang}}, \bibinfo {author} {\bibfnamefont {M.}~\bibnamefont {Sasaki}}, \bibinfo {author} {\bibfnamefont {A.}~\bibnamefont {Sinclair}}, \emph {et~al.},\ }in\ \href@noop {} {\emph {\bibinfo {booktitle} {Advanced Photon Counting Techniques IV}}},\ Vol.\ \bibinfo {volume} {7681}\ (\bibinfo {organization} {SPIE},\ \bibinfo {year} {2010})\ pp.\ \bibinfo {pages} {81--91}\BibitemShut {NoStop}%
\bibitem [{\citenamefont {Yankowitz}\ \emph {et~al.}(2019)\citenamefont {Yankowitz}, \citenamefont {Ma}, \citenamefont {Jarillo-Herrero},\ and\ \citenamefont {LeRoy}}]{yankowitz2019van}%
  \BibitemOpen
  \bibfield  {author} {\bibinfo {author} {\bibfnamefont {M.}~\bibnamefont {Yankowitz}}, \bibinfo {author} {\bibfnamefont {Q.}~\bibnamefont {Ma}}, \bibinfo {author} {\bibfnamefont {P.}~\bibnamefont {Jarillo-Herrero}},\ and\ \bibinfo {author} {\bibfnamefont {B.~J.}\ \bibnamefont {LeRoy}},\ }\href@noop {} {\bibfield  {journal} {\bibinfo  {journal} {Nature Reviews Physics}\ }\textbf {\bibinfo {volume} {1}},\ \bibinfo {pages} {112} (\bibinfo {year} {2019})}\BibitemShut {NoStop}%
\bibitem [{\citenamefont {Daveau}\ \emph {et~al.}(2020)\citenamefont {Daveau}, \citenamefont {Vandekerckhove}, \citenamefont {Mukherjee}, \citenamefont {Wang}, \citenamefont {Shan}, \citenamefont {Mak}, \citenamefont {Vamivakas},\ and\ \citenamefont {Fuchs}}]{daveau2020spectral}%
  \BibitemOpen
  \bibfield  {author} {\bibinfo {author} {\bibfnamefont {R.~S.}\ \bibnamefont {Daveau}}, \bibinfo {author} {\bibfnamefont {T.}~\bibnamefont {Vandekerckhove}}, \bibinfo {author} {\bibfnamefont {A.}~\bibnamefont {Mukherjee}}, \bibinfo {author} {\bibfnamefont {Z.}~\bibnamefont {Wang}}, \bibinfo {author} {\bibfnamefont {J.}~\bibnamefont {Shan}}, \bibinfo {author} {\bibfnamefont {K.~F.}\ \bibnamefont {Mak}}, \bibinfo {author} {\bibfnamefont {A.~N.}\ \bibnamefont {Vamivakas}},\ and\ \bibinfo {author} {\bibfnamefont {G.~D.}\ \bibnamefont {Fuchs}},\ }\href@noop {} {\bibfield  {journal} {\bibinfo  {journal} {Apl Photonics}\ }\textbf {\bibinfo {volume} {5}} (\bibinfo {year} {2020})}\BibitemShut {NoStop}%
\bibitem [{\citenamefont {Piccinini}\ \emph {et~al.}(2024)\citenamefont {Piccinini}, \citenamefont {Paralikis}, \citenamefont {Neto}, \citenamefont {Madigawa}, \citenamefont {Wyborski}, \citenamefont {Remesh}, \citenamefont {Vannucci}, \citenamefont {Gregersen}, \citenamefont {Munkhbat} \emph {et~al.}}]{piccinini2024high}%
  \BibitemOpen
  \bibfield  {author} {\bibinfo {author} {\bibfnamefont {C.}~\bibnamefont {Piccinini}}, \bibinfo {author} {\bibfnamefont {A.}~\bibnamefont {Paralikis}}, \bibinfo {author} {\bibfnamefont {J.}~\bibnamefont {Neto}}, \bibinfo {author} {\bibfnamefont {A.~A.}\ \bibnamefont {Madigawa}}, \bibinfo {author} {\bibfnamefont {P.}~\bibnamefont {Wyborski}}, \bibinfo {author} {\bibfnamefont {V.}~\bibnamefont {Remesh}}, \bibinfo {author} {\bibfnamefont {L.}~\bibnamefont {Vannucci}}, \bibinfo {author} {\bibfnamefont {N.}~\bibnamefont {Gregersen}}, \bibinfo {author} {\bibfnamefont {B.}~\bibnamefont {Munkhbat}}, \emph {et~al.},\ }\href@noop {} {\bibfield  {journal} {\bibinfo  {journal} {arXiv preprint arXiv:2406.07097}\ } (\bibinfo {year} {2024})}\BibitemShut {NoStop}%
\bibitem [{\citenamefont {Kramer}(2023)}]{kramer2023helium}%
  \BibitemOpen
  \bibfield  {author} {\bibinfo {author} {\bibfnamefont {D.}~\bibnamefont {Kramer}},\ }\href@noop {} {\bibfield  {journal} {\bibinfo  {journal} {Physics Today}\ }\textbf {\bibinfo {volume} {76}},\ \bibinfo {pages} {18} (\bibinfo {year} {2023})}\BibitemShut {NoStop}%
\bibitem [{\citenamefont {Gemmell}\ \emph {et~al.}(2017)\citenamefont {Gemmell}, \citenamefont {Hills}, \citenamefont {Bradshaw}, \citenamefont {Rawlings}, \citenamefont {Green}, \citenamefont {Heath}, \citenamefont {Tsimvrakidis}, \citenamefont {Dobrovolskiy}, \citenamefont {Zwiller}, \citenamefont {Dorenbos} \emph {et~al.}}]{gemmell2017miniaturized}%
  \BibitemOpen
  \bibfield  {author} {\bibinfo {author} {\bibfnamefont {N.~R.}\ \bibnamefont {Gemmell}}, \bibinfo {author} {\bibfnamefont {M.}~\bibnamefont {Hills}}, \bibinfo {author} {\bibfnamefont {T.}~\bibnamefont {Bradshaw}}, \bibinfo {author} {\bibfnamefont {T.}~\bibnamefont {Rawlings}}, \bibinfo {author} {\bibfnamefont {B.}~\bibnamefont {Green}}, \bibinfo {author} {\bibfnamefont {R.~M.}\ \bibnamefont {Heath}}, \bibinfo {author} {\bibfnamefont {K.}~\bibnamefont {Tsimvrakidis}}, \bibinfo {author} {\bibfnamefont {S.}~\bibnamefont {Dobrovolskiy}}, \bibinfo {author} {\bibfnamefont {V.}~\bibnamefont {Zwiller}}, \bibinfo {author} {\bibfnamefont {S.~N.}\ \bibnamefont {Dorenbos}}, \emph {et~al.},\ }\href@noop {} {\bibfield  {journal} {\bibinfo  {journal} {Superconductor Science and Technology}\ }\textbf {\bibinfo {volume} {30}},\ \bibinfo {pages} {11LT01} (\bibinfo {year} {2017})}\BibitemShut {NoStop}%
\bibitem [{\citenamefont {Zugliani}\ \emph {et~al.}(2023)\citenamefont {Zugliani}, \citenamefont {Schmid}, \citenamefont {Wietschorke}, \citenamefont {Jonas}, \citenamefont {Spedicato}, \citenamefont {Strohauer}, \citenamefont {Grotowski}, \citenamefont {Flaschmann}, \citenamefont {M{\"u}ller}, \citenamefont {Althammer} \emph {et~al.}}]{zugliani2023tailoring}%
  \BibitemOpen
  \bibfield  {author} {\bibinfo {author} {\bibfnamefont {L.}~\bibnamefont {Zugliani}}, \bibinfo {author} {\bibfnamefont {C.}~\bibnamefont {Schmid}}, \bibinfo {author} {\bibfnamefont {F.}~\bibnamefont {Wietschorke}}, \bibinfo {author} {\bibfnamefont {B.}~\bibnamefont {Jonas}}, \bibinfo {author} {\bibfnamefont {S.}~\bibnamefont {Spedicato}}, \bibinfo {author} {\bibfnamefont {S.}~\bibnamefont {Strohauer}}, \bibinfo {author} {\bibfnamefont {S.}~\bibnamefont {Grotowski}}, \bibinfo {author} {\bibfnamefont {R.}~\bibnamefont {Flaschmann}}, \bibinfo {author} {\bibfnamefont {M.}~\bibnamefont {M{\"u}ller}}, \bibinfo {author} {\bibfnamefont {M.}~\bibnamefont {Althammer}}, \emph {et~al.},\ }in\ \href@noop {} {\emph {\bibinfo {booktitle} {2023 IEEE Globecom Workshops (GC Wkshps)}}}\ (\bibinfo {organization} {IEEE},\ \bibinfo {year} {2023})\ pp.\ \bibinfo {pages} {1051--1056}\BibitemShut {NoStop}%
\bibitem [{\citenamefont {Dauler}\ \emph {et~al.}(2014)\citenamefont {Dauler}, \citenamefont {Grein}, \citenamefont {Kerman}, \citenamefont {Marsili}, \citenamefont {Miki}, \citenamefont {Nam}, \citenamefont {Shaw}, \citenamefont {Terai}, \citenamefont {Verma},\ and\ \citenamefont {Yamashita}}]{dauler2014review}%
  \BibitemOpen
  \bibfield  {author} {\bibinfo {author} {\bibfnamefont {E.~A.}\ \bibnamefont {Dauler}}, \bibinfo {author} {\bibfnamefont {M.~E.}\ \bibnamefont {Grein}}, \bibinfo {author} {\bibfnamefont {A.~J.}\ \bibnamefont {Kerman}}, \bibinfo {author} {\bibfnamefont {F.}~\bibnamefont {Marsili}}, \bibinfo {author} {\bibfnamefont {S.}~\bibnamefont {Miki}}, \bibinfo {author} {\bibfnamefont {S.~W.}\ \bibnamefont {Nam}}, \bibinfo {author} {\bibfnamefont {M.~D.}\ \bibnamefont {Shaw}}, \bibinfo {author} {\bibfnamefont {H.}~\bibnamefont {Terai}}, \bibinfo {author} {\bibfnamefont {V.~B.}\ \bibnamefont {Verma}},\ and\ \bibinfo {author} {\bibfnamefont {T.}~\bibnamefont {Yamashita}},\ }\href@noop {} {\bibfield  {journal} {\bibinfo  {journal} {Optical Engineering}\ }\textbf {\bibinfo {volume} {53}},\ \bibinfo {pages} {081907} (\bibinfo {year} {2014})}\BibitemShut {NoStop}%
\end{thebibliography}%

\end{document}